\documentclass[twocolumn]{aastex631}

\newcommand{\xmm}{{\it XMM-Newton}}

\newcommand{\fluxcgs}{erg\,cm$^{-2}$\,s$^{-1}$}

\newcommand{\es}{{1ES~1959+650}}
\newcommand{\xrt}{{\it Swift}-XRT}
\newcommand{\ixpe}{{\it IXPE}}
\newcommand{\one}{May 3-4}
\newcommand{\two}{June 9-12}
\newcommand\rev{}

\shorttitle{X-ray Polarization of \es}
\shortauthors{Errando et al.}

\begin{document}

\title{Detection of X-ray Polarization from the Blazar \es\ with the Imaging X-ray Polarimetry Explorer}

% % LEADS
\author[0000-0002-1853-863X]{Manel Errando}
\correspondingauthor{Manel Errando}
\email{errando@wustl.edu}
\affiliation{Physics Department and McDonnell Center for the Space Sciences, Washington University in St. Louis,
St. Louis, MO 63130, USA}
\author[0000-0001-9200-4006]{Ioannis Liodakis}\affiliation{Finnish Centre for Astronomy with ESO, 20014 University of Turku, Finland}\affiliation{NASA Marshall Space Flight Center, Huntsville, AL 35812, USA}
\author[0000-0001-7396-3332]{Alan P. Marscher}\affiliation{Institute for Astrophysical Research, Boston University, 725 Commonwealth Avenue, Boston, MA 02215, USA}

% TIER-1

\author[0000-0002-6492-1293]{Herman L. Marshall}\affiliation{MIT Kavli Institute for Astrophysics and Space Research, Massachusetts Institute of Technology, 77 Massachusetts Avenue, Cambridge, MA 02139, USA}
\author{Riccardo Middei}
\affiliation{Space Science Data Center, Agenzia Spaziale Italiana, Via del Politecnico snc, 00133 Roma, Italy}
\affiliation{INAF Osservatorio Astronomico di Roma, Via Frascati 33, 00078 Monte Porzio Catone (RM), Italy}
\author[0000-0002-6548-5622]{Michela Negro}\affiliation{Department of Physics and Astronomy, Louisiana State University, Baton Rouge, LA 70803, USA}%\affiliation{NASA Goddard Space Flight Center, Greenbelt, MD 20771, USA}\affiliation{Center for Research and Exploration in Space Science and Technology, NASA/GSFC, Greenbelt, MD 20771, USA}
\author[0000-0001-6292-1911]{Abel Lawrence Peirson}\affiliation{Department of Physics and Kavli Institute for Particle Astrophysics and Cosmology, Stanford University, Stanford, California 94305, USA}
\author[0000-0003-3613-4409]{Matteo Perri}\affiliation{Space Science Data Center, Agenzia Spaziale Italiana, Via del Politecnico snc, 00133 Roma, Italy}\affiliation{INAF Osservatorio Astronomico di Roma, Via Frascati 33, 00078 Monte Porzio Catone (RM), Italy}
\author[0000-0002-2734-7835]{Simonetta Puccetti}\affiliation{Space Science Data Center, Agenzia Spaziale Italiana, Via del Politecnico snc, 00133 Roma, Italy}
\author[0000-0002-5104-5263]{Pazit L. Rabinowitz}
\affiliation{Physics Department and McDonnell Center for the Space Sciences, Washington University in St. Louis,
St. Louis, MO 63130, USA}

% MW partners

% ...

\begin{abstract}

Observations of linear polarization in the 2-8\,keV energy range with the Imaging X-ray Polarimetry Explorer (\ixpe) explore the magnetic field geometry and dynamics of the regions generating non-thermal radiation in relativistic jets of blazars.
These jets, particularly in blazars whose spectral energy distribution peaks at X-ray energies, emit X-rays via synchrotron radiation from high-energy particles within the jet. \ixpe\ observations 
of the X-ray selected BL Lac-type blazar \es\ in 2022 \one\ showed a 
significant linear polarization 
degree of $\Pi_\mathrm{x} = 8.0\% \pm 2.3\%$ at an electric-vector position angle $\psi_\mathrm{x} = 123^\circ \pm 8^\circ$. However, in 2022 \two, only an upper limit of $\Pi_\mathrm{x} \leq 5.1\%$ could be derived (at the 99\% confidence level). The degree of optical polarization at that time $\Pi_\mathrm{O} \sim 5\%$ is comparable to the X-ray measurement. We investigate possible scenarios for these findings, 
including temporal and geometrical depolarization effects. Unlike some other X-ray selected BL~Lac objects, there is no significant chromatic dependence of the measured polarization in \es, and its low X-ray polarization may be attributed to 
turbulence in the jet flow with dynamical timescales shorter than 1 day.

\end{abstract}


\section{Introduction} \label{sec:intro}

Relativistic jets are powerful streams of collimated plasma and radiation that play a prominent role in various astrophysical phenomena, such as active galactic nuclei, gamma-ray bursts, and X-ray binary systems 
{\rev \citep{1984RvMP...56..255B,2004A&A...414..895F, Hughes2006,2019ARA&A..57..467B}}. In the case of blazars, these jets are fueled by accretion onto a central supermassive black hole within an active galactic nucleus, and are oriented in a direction that is closely aligned with Earth's line of sight \citep{1995PASP..107..803U}. These jets accelerate particles to energies beyond $10^{10}$\,eV, producing non-thermal emission observed across a wide range of frequencies, from radio to very high-energy (VHE, $>0.1$  TeV) $\gamma$ rays \citep{2012A&A...541A.160G,Liodakis2019}. 
Advancements in understanding the physics of relativistic jets rely heavily on multi-wavelength observations encompassing the spectral energy distribution, flux variability, and polarization of the observed emission. 

According to most theoretical models for the production, acceleration, and collimation of relativistic jets, the plasma is Poynting-flux dominated close to the black hole, where the jet is accelerated and collimated \citep[e.g.,][]{Blandford1977,Vlahakis2004}. Near the end of the acceleration and collimation zone, there is a transition to a particle-dominated flow \citep[e.g.,][]{Lyubarsky2010}. It is currently unclear, however, where most of the particle acceleration and radiative dissipation take place. Multiple processes could play major roles in energizing the particles that produce the radiation in blazar jets. These include magnetic reconnection events \citep[e.g.,][]{1992A&A...262...26R,2015SSRv..191..545K,2018MNRAS.473.4840W}, relativistic shocks \citep[e.g.,][]{1978ApJ...221L..29B,2003ApJ...595..555N,2015SSRv..191..519S}, {\rev and stochastic particle acceleration \citep[e.g.,][]{1996ApJ...456..106D,1996A&A...314.1010K,2006A&A...453...47K}. These} mechanisms can efficiently accelerate particles, although particle-in-cell simulations find that magnetic reconnection is more efficient than shocks at doing so where the magnetization level is high \citep{2014ApJ...783L..21S}. On the other hand, shocks can be more efficient in regions where the magnetization is low \citep{Sironi2015}.

Multi-wavelength observations of linear polarization probe the geometry of the magnetic field in different locations in relativistic jets \citep[e.g.,][]{1980ARA&A..18..321A,Jorstad2005AJ....130.1418J,marscher2021}. The results can indicate which of the particle acceleration mechanisms are operating in sites of strong high-energy radiation \citep[e.g.,][]{zhang2019,tavecchio2021,zhang2022,DiGesu2022models,marscher2022}. Such polarization observations of blazars have long been available at radio, infrared and optical wavelengths.
Observations in the X-ray band now benefit from the availability of linear polarization sensitivity through the 
Imaging X-ray Polarimetry Explorer satellite \citep[\ixpe,][]{weisskopf2022}, which began
science operations on 2022 January 11. 
\ixpe\ consists of three 4\,m focal-length X-ray mirrors that focus on three identical polarization-sensitive gas pixel detector units (DU1-3). The sensitivity of an X-ray polarimeter is commonly assessed by its minimum detectable polarization at 99\% confidence level, $MDP_{99}$. At a flux level of $F_{2-8\,\mathrm{keV}} = 10^{-10}$\,\fluxcgs, \ixpe\ achieves $MDP_{99} \approx 8\%$ in a 50\,ks exposure. 

The BL Lacertae object \es\ \citep[$z=0.047$, ][]{1991rc3..book.....D} stands out as one of the brightest X-ray blazars. Only four BL~Lac objects in the Einstein Slew Survey exhibit a higher X-ray flux \citep{1996ApJS..104..251P}. It is also among the first blazars from which TeV $\gamma$ rays have been detected \citep{2003ApJ...583L...9H}. 
The X-ray flux of \es\ generally exceeds $10^{-10}$\fluxcgs\ in the 0.1-10 keV range, displaying variability on timescales $<1$\,h \citep[][]{Kapanadze2014}. 
While X-ray and $\gamma$-ray fluxes are generally found to be correlated, a notable ``orphan'' TeV flare without an X-ray counterpart was detected by the Whipple observatory in 2002 \citep{2004ApJ...601..151K}. Owing to its bright X-ray emission, \es\ was among the first blazars observed by \ixpe. 

The radio to X-ray emission from \es\ is commonly interpreted as originating from synchrotron radiation by relativistic electrons cooling in the jet's magnetic field. Synchrotron radiation is inherently polarized. {\rev The expected degree of polarization from a power-law energy spectrum of 
relativistic electrons $N\left(E_e\right)\propto E_e^{-s}$ in a homogeneous magnetic field is $\Pi = \left(s+1\right)/\left(s+7/3\right) \approx 70\%$ for $s \approx 2$ \citep{1979rpa..book.....R}.} %$\Pi = \left(s+1\right)/\left(s+7/3\right)$. 
If the magnetic field configuration has a random component $B_r$ then the polarization degree will be suppressed by a factor $B_0^2/(B_0^2 + B_r^2)$, where $B_0$ is the intensity of the ordered magnetic field component \citep{1965ARA&A...3..297G}. In blazars, the presence of a random component in the magnetic field reduces the polarization signal from a maximum value of order 75\% to an average value that is usually between a few and tens of percent \citep[e.g.,][]{Blinov2021,marscher2021}. Therefore, polarization measurements in the X-ray band by \ixpe\ are sensitive to the relative ratio of the random to the ordered magnetic field in the region of the jet where the non-thermal emission from the highest-energy particles is emitted. This new probe of the magnetic field intensity and geometry in relativistic jets can be used to discern among the possible models for particle acceleration. 
For instance, if acceleration occurs in stationary shocks, downstream of which particles lose energy, we anticipate steady X-ray emission with higher polarization than than the polarization of the optical emission by lower-energy particles radiating over a larger volume \citep{Angelakis2016, tavecchio2018, tavecchio2020, zhang2022, DiGesu2022models, marscher2022}. Conversely, magnetic reconnection events convert magnetic energy into kinetic energy, resulting in moderate X-ray and optical polarization, with a weaker dependence on photon energy \citep[e.g.,][]{Bodo2021, zhang2022}.

Here we present the first detection of linear X-ray polarization from \es\ by \ixpe\ in 2022 \one\ and \two. We describe the \ixpe\ measurements, as well as X-ray spectral observations with the Neil Gehrels Swift X-ray Telescope (\xrt) and \xmm\
in \S {\ref{sec:xray}}, and observations at other wavebands in \S 
 {\ref{sec:mw}}. We discuss and interpret the results in \S\ref{sec:discussion} and summarize our findings in \S\ref{sec:conclusions}.

\begin{table*}[ht]
\centering
\caption{Summary and results from \ixpe, \xmm, and \xrt\ X-ray observations in 2022 \one\ and \two.}
\begin{tabular}{lcclcccc}
\hline
Telescope & Energy range & ObsID & Date &  Exposure & $F_{2-8\,\mathrm{keV}}$ & $\Gamma_x$ & $\chi^2/dof$\\
 & [keV] & & & [ks] & [$10^{-10}$\,erg\,cm$^{-2}$\,s$^{-1}$] & &\\
\hline
\hline
2022 \one \\
\hline
\ixpe\    & 2.0-8.0  &  01006201    & 2022 05 03-04 & 54 & $1.24\pm 0.02$ & $2.50 \pm 0.02$ & $30.6/26$ \\
\xrt\     & 0.5-10   &  00096560006 & 2022 05 03    & 0.9   & $1.57 \pm 0.04$ & $2.18 \pm 0.04$  & $314/349$\\ 
\xrt\     & 0.5-10   &  00096560007 & 2022 05 04    & 0.9  & $1.63 \pm 0.05$ & $2.17 \pm 0.04$  & $374/351$ \\ 
\xmm\     &   1.0-10  &  0902110801  & 2022 05 06    & 16  & $1.35 \pm 0.01$ & $2.26 \pm 0.01$  & $196/155$\\ 
\hline
2022 \two \\
\hline
\ixpe\    & 2.0-8.0  &  01006001    & 2022 06 09-12 & 200  & $1.47 \pm 0.01$ & $2.29 \pm 0.01$ & $36.5/25$\\ 
\xrt\     & 0.5-10   &  00096560012 & 2022 06 12    & 0.9   & $2.23 \pm 0.05$ & $2.20 \pm 0.04$   & $334/344$\\ 
\xmm      &  1.0-10   &  0902111201  & 2022 06 23    & 18  & $1.34 \pm 0.01 $ & $2.20 \pm 0.01 $ & $195/155$ \\ 
\hline
\end{tabular}
\label{tab:obs}
\end{table*}

\section{X-ray observations} \label{sec:xray}
\ixpe\ first observed \es\ on 2022 May 3-4 with an exposure time of 54\,ks. 
A second science observation was carried out on 2022 June 9-12, accumulating 200\,ks of exposure. \xmm\ \citep{2001A&A...365L..18S} collected two contemporaneous  exposures of \es, and 
\xrt\ \citep{2005SSRv..120..165B} monitored the source with a total of eleven exposures during the duration of campaign. 
\begin{figure*}[ht]
\includegraphics[height=3in]{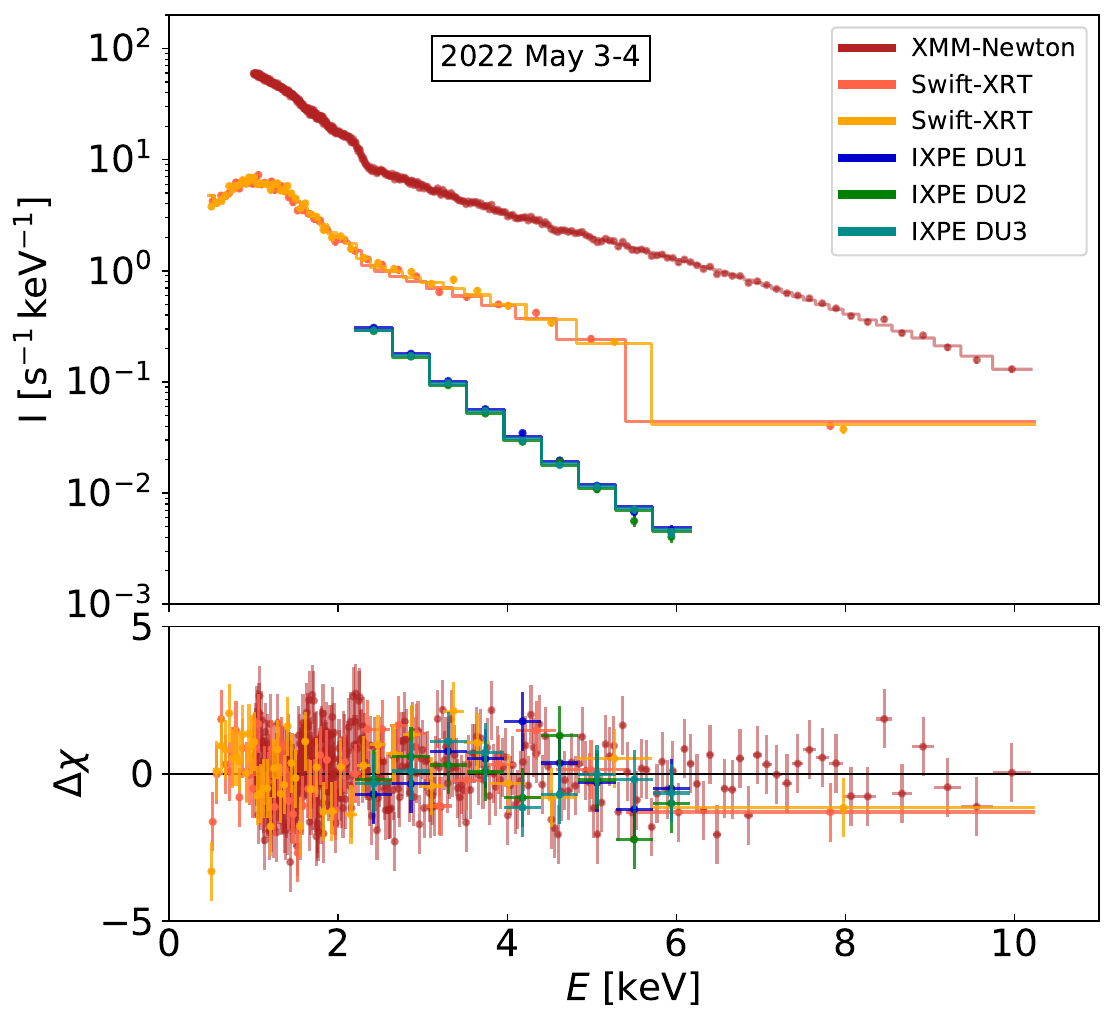}
\hfill
\includegraphics[height=3in]{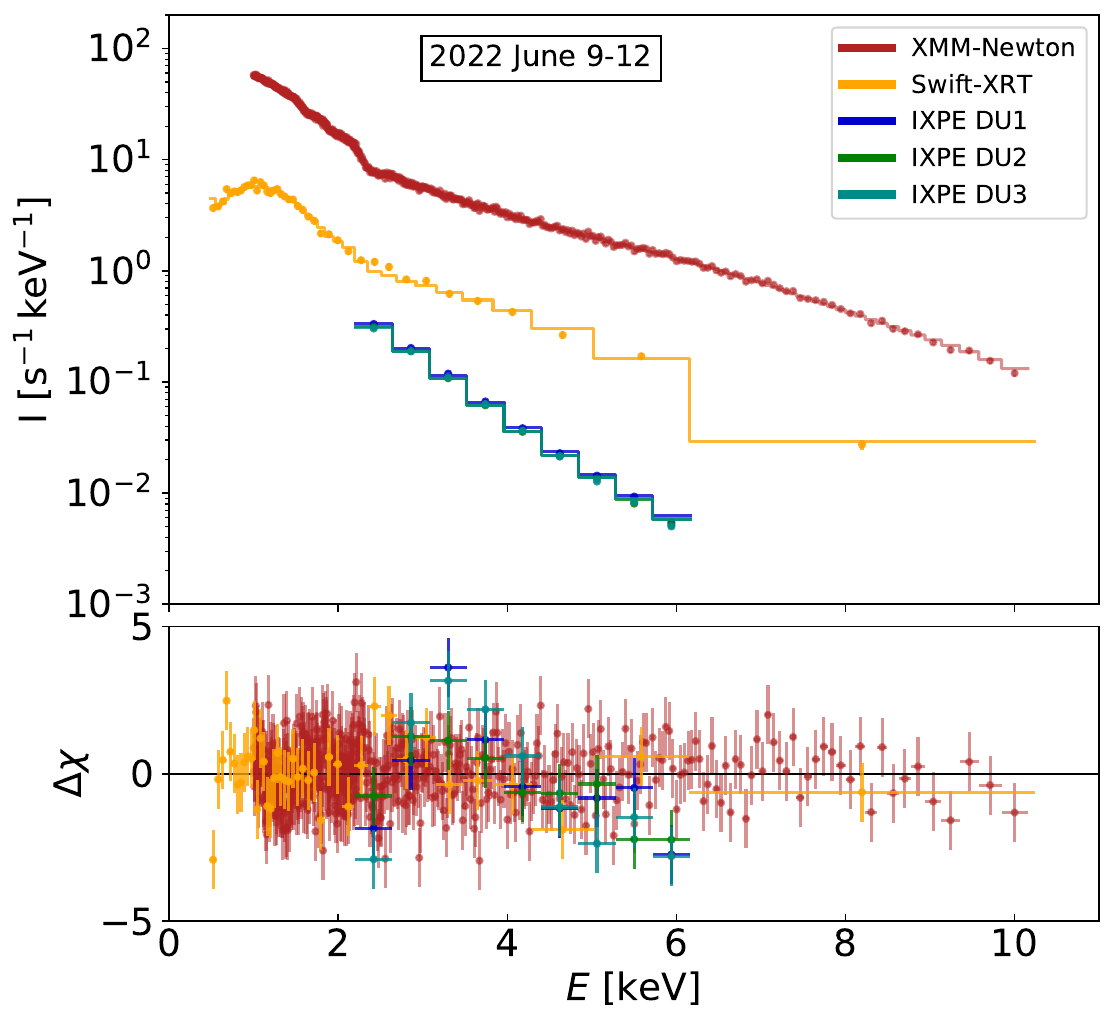} 
 \caption{Spectral fit to the X-ray data of \es\ during \ixpe\ observations in 2022 \one\ (\textit{left panel}) and 2022 \two\ (\textit{right panel}). The \ixpe\ Stokes $I$, \xrt, and \xmm\ spectra are fit to derive the column density absorption and determine the presence of spectral curvature. These results will be used in the later spectro-polarimetric fit of the \ixpe\ $I\,,Q\,,\mathrm{and}\ U$ spectra.}
\label{fig:xspec}
\end{figure*}

\begin{figure*}
\centering
\includegraphics[width=0.85\textwidth]{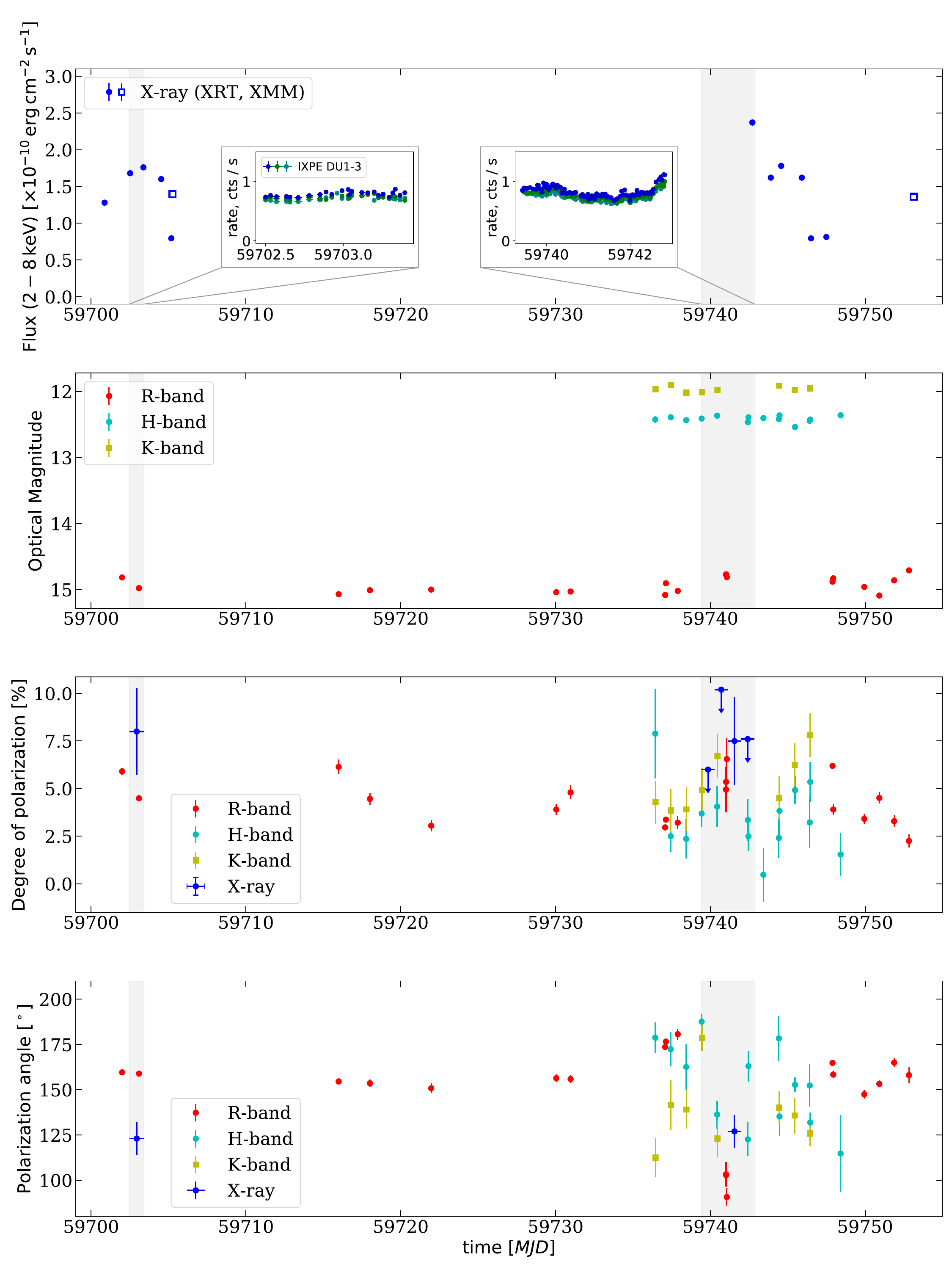}
 \caption{X-ray, optical, and IR light curves contemporaneous with the \ixpe\ observations of \es. Observations are described in \S\ref{sec:xray} and \S\ref{sec:mw}. Panels show, from top to bottom, X-ray flux, optical and IR brightness in magnitudes, degree of polarization, and polarization angle. Inset plots in the top panel display the X-ray light curve measured by \ixpe. {\rev Blue data points represent significant measurements of X-ray polarization, while blue downward-pointing arrows indicate 99\% confidence level upper limits during time periods without a significant detection X-ray polarization. }Grey vertical shaded areas indicate the two epochs of observation of \ixpe. } 
\label{fig:pol_light}
\end{figure*}

We combine the \ixpe\ Stokes $I$, \xrt, and \xmm\ spectra to {\rev characterize the energy spectrum of \es\ in the X-ray band. The goal of this combined fit is to}
determine the column density absorption and spectral shape of the X-ray spectrum of \es, as the 2-8\,keV energy range of \ixpe\ is too narrow to robustly measure the level of neutral density absorption and the potential presence of spectral curvature.  
The flux normalizations and spectral properties are left free to allow for inter-calibration uncertainties \citep{2017AJ....153....2M} as well as flux and spectral variability on timescales of days, which is often observed in \es\ even in relatively quiescent X-ray states \citep{2008ApJ...679.1029T}. 
The fit of the combined data set to an absorbed power law $dN/dE \propto E^{-\Gamma_\mathrm{x}}$ yields a $\chi^2$/dof fit statistic of {\rev $554/497$} (with associated probability {\rev $p=0.04$}) for the \one\ data and $616/566$ ($p=0.07$) for \two\ (Figure~\ref{fig:xspec}). The best-fit parameters describing the spectra are listed in Table~\ref{tab:obs}. 
Changing the model to an absorbed log-parabola does not significantly improve the fit in either case. The derived column density of $(1.32 \pm 0.04) \times 10^{21}$\,cm$^{-2}$, exceeding the Galactic value by 30\% \citep{2016A&A...594A.116H}, suggests the presence of additional neutral absorption along the line of sight within the host galaxy to the object, consistent with previous studies \citep{2013ApJ...775....3A,2014ApJ...797...89A}. 
During both epochs, the {\rev average} X-ray flux is within 10\% of the median value measured by \xrt\ between 2005 and 2022 \citep[Figure~\ref{fig:pol_light};][]{Falcone2013ApJS..207...28S}. We conclude that \es\ was in an average X-ray flux state during the \ixpe\ campaign. 

\begin{figure*}[ht]
\resizebox{\hsize}{!}{ \includegraphics[height=0.96in]{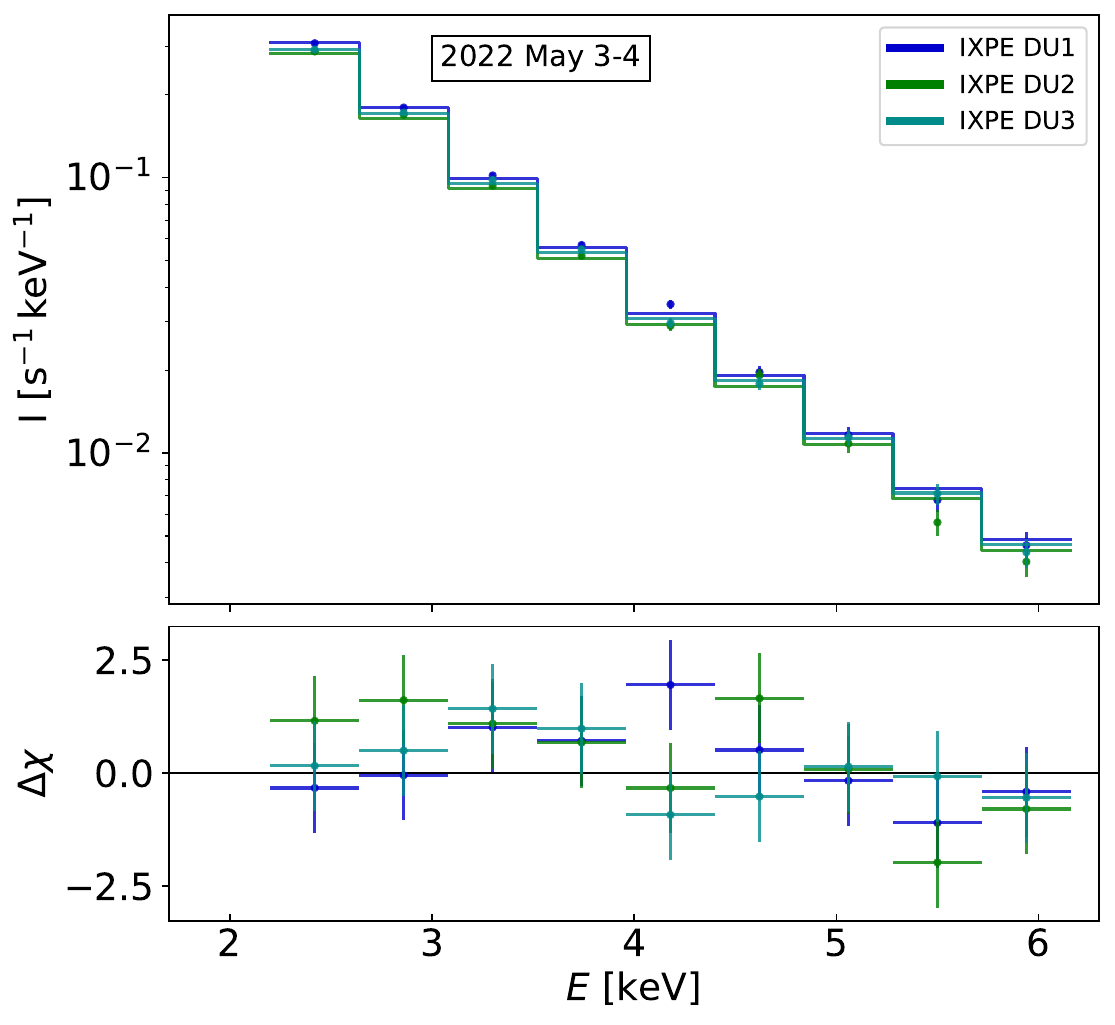}     \includegraphics[height=1in]{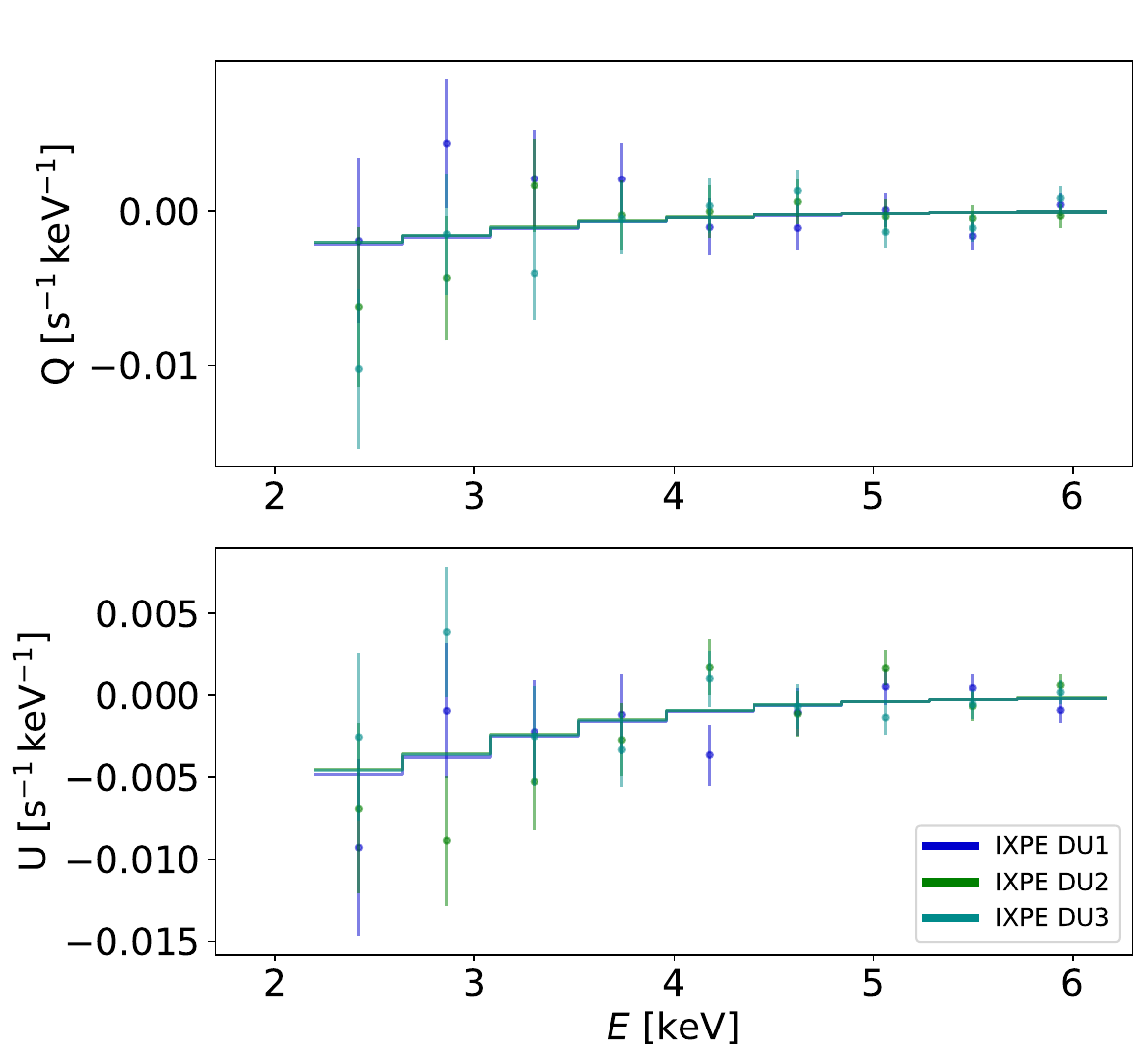}}
\resizebox{\hsize}{!}{ \includegraphics[height=0.96in]{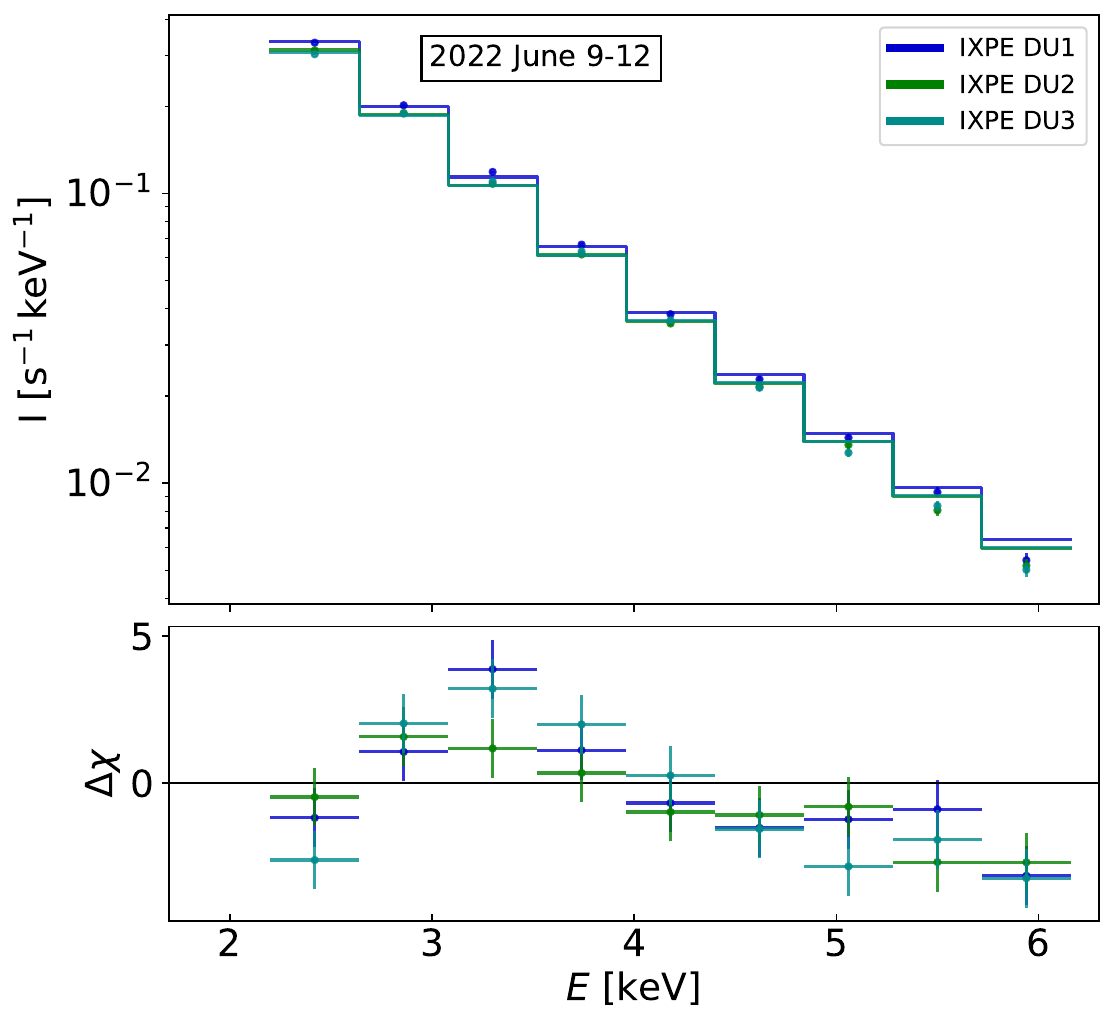}     \includegraphics[height=1in]{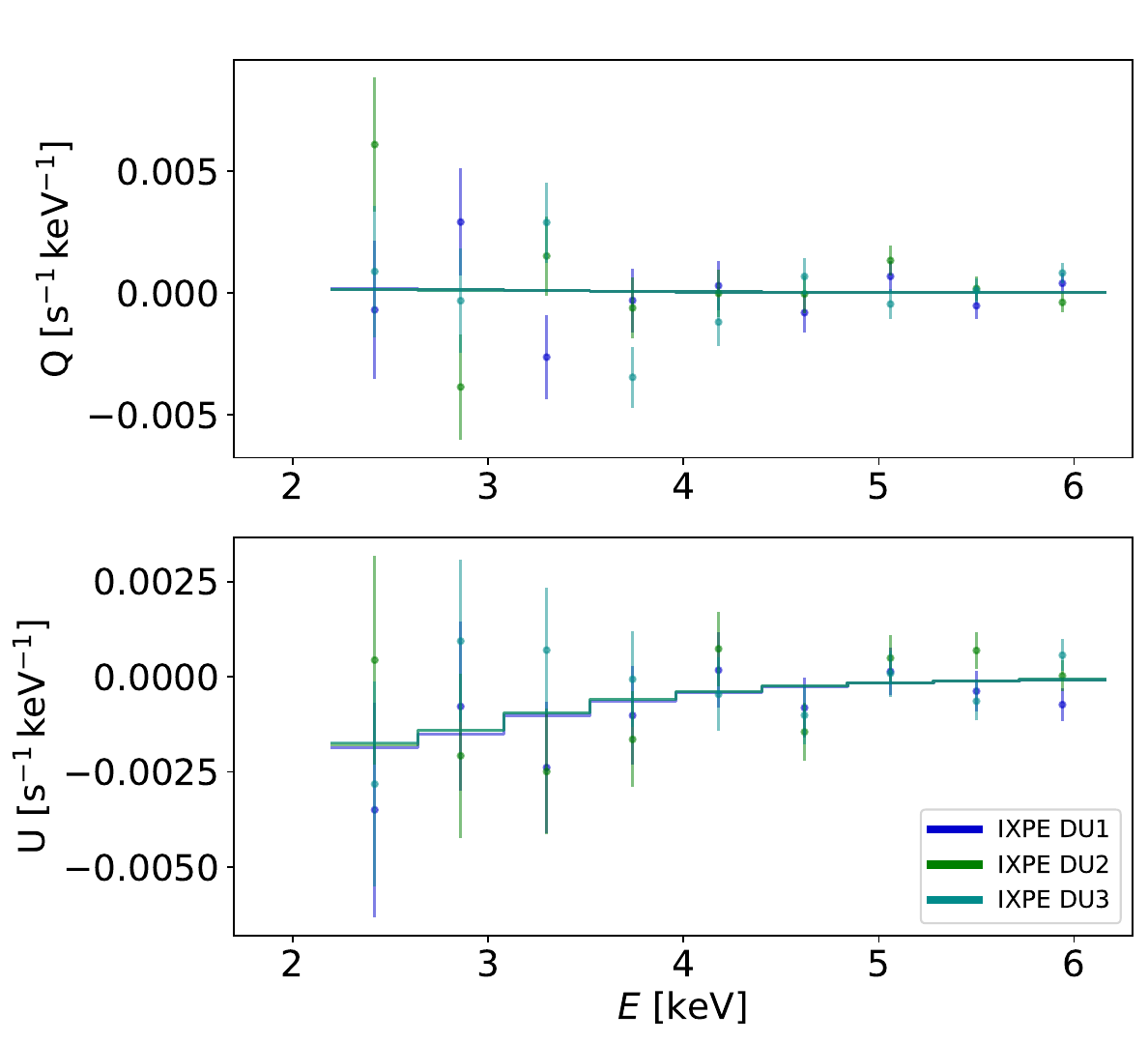}}
 \caption{Spectro-polarimetric fit to the X-ray data of \es\ during \ixpe\ observations in 2022 \one\ (\textit{top panels}) and 2022 \two\ (\textit{bottom panels}). Panels on the left present the  fit to the \ixpe\ Stokes $I$ with its residuals, while panels on the right show the fits to the \ixpe\ $Q$ and $U$ spectra. }
\label{fig:pol_obs}
\end{figure*}

We have searched for time-averaged X-ray polarization from \es\ in the \ixpe\ data. First the $I$ spectra were fit with an absorbed power-law model. The absorbing column density was fixed to the value obtained during the combined spectral fit, but the powerlaw normalization and photon index were allowed to vary. We then tested for the presence of a constant degree of polarization and angle as a function of energy by performing a spectropolarimetric fit to the Stokes $I$, $Q$, and $U$ spectra, with the only free parameters being the degree of polarization and polarization angle (Figure~\ref{fig:pol_obs}). 

In \one, the spectrum obtained from \ixpe\ displayed a softer photon index of $\Gamma = 2.50 \pm 0.02$ compared to the contemporaneous \xmm\ and \xrt\ data (Table~\ref{tab:obs}). In \two\ the \ixpe-derived $\Gamma = 2.29 \pm 0.01$ was in closer agreement with the value previously derived. The effect is likely to be the result of an improvement in the telescope alignment between the two  observations. The polarization degree in \one\  was $\Pi_\mathrm{x} = 8.0\% \pm 2.3\%$, with an electric-vector position angle $\psi_\mathrm{x} = 123\degr \pm 8\degr$. A null hypothesis of no linear polarization can be excluded at $3.5\,\sigma$ confidence level. No significant linear polarization was detected in \two. From this second \ixpe\ observation, we derive an  upper limit of $\Pi_\mathrm{x} < 5.1\%$ (99\% confidence level), with the polarization angle left unconstrained. The hypothesis that the degree of polarization remained constant across the \one\ and \two\ \ixpe\ observations has an associated probability of $P=0.0244$ using a $\chi^2$ test, i.e. it can only be excluded at $2.3\,\sigma$ level.

To search for time-dependent X-ray polarization, we split the 200\,ks-duration \ixpe\ exposure from \two\ into four $50$\,ks slices and repeated the spectro-polarimetric fit described above independently for each time bin. We found no significant linear polarization in three out of the four segments, with only the third one showing significant polarization corresponding to a $2.8\,\sigma$ post-trial confidence level (Figure~\ref{fig:pol_light}). The constraints on the degree of polarization in the four time bins are $\Pi_\mathrm{x}  < 6.0\%,\,<10.2\%,\,7.5\% \pm 2.3\%,\, \mathrm{and} <7.6\%$ (upper limits are 99\% confidence level). The polarization angle is only constrained for the third time bin to be $\psi_\mathrm{x} = 127^\circ \pm 9^\circ$.

X-ray data reduction procedures are detailed in Appendix~\ref{app:x}.

\section{Radio, infrared, and optical observations}
\label{sec:mw}

Figure~\ref{fig:pol_light} summarizes the time evolution on the X-ray and optical polarization measurements of \es. Contemporaneous polarization observations were obtained at radio (millimeter), infrared (IR), and optical wavelengths using the Institut de Radioastronomie Millim\'etrique (IRAM) 30\,m radio telescope, the 1.8\,m Perkins Telescope, the Calar alto observatory, the KANATA telescope, the Nordic Optical Telescope (NOT), the Sierra Nevada Observatory, RoboPol mounted at the 1.3 m telescope of the Skinakas Observatory, {the Submillimeter Array (SMA),} and the Very Long Baseline Array (VLBA). 

\begin{figure*}[ht]%
\plotone{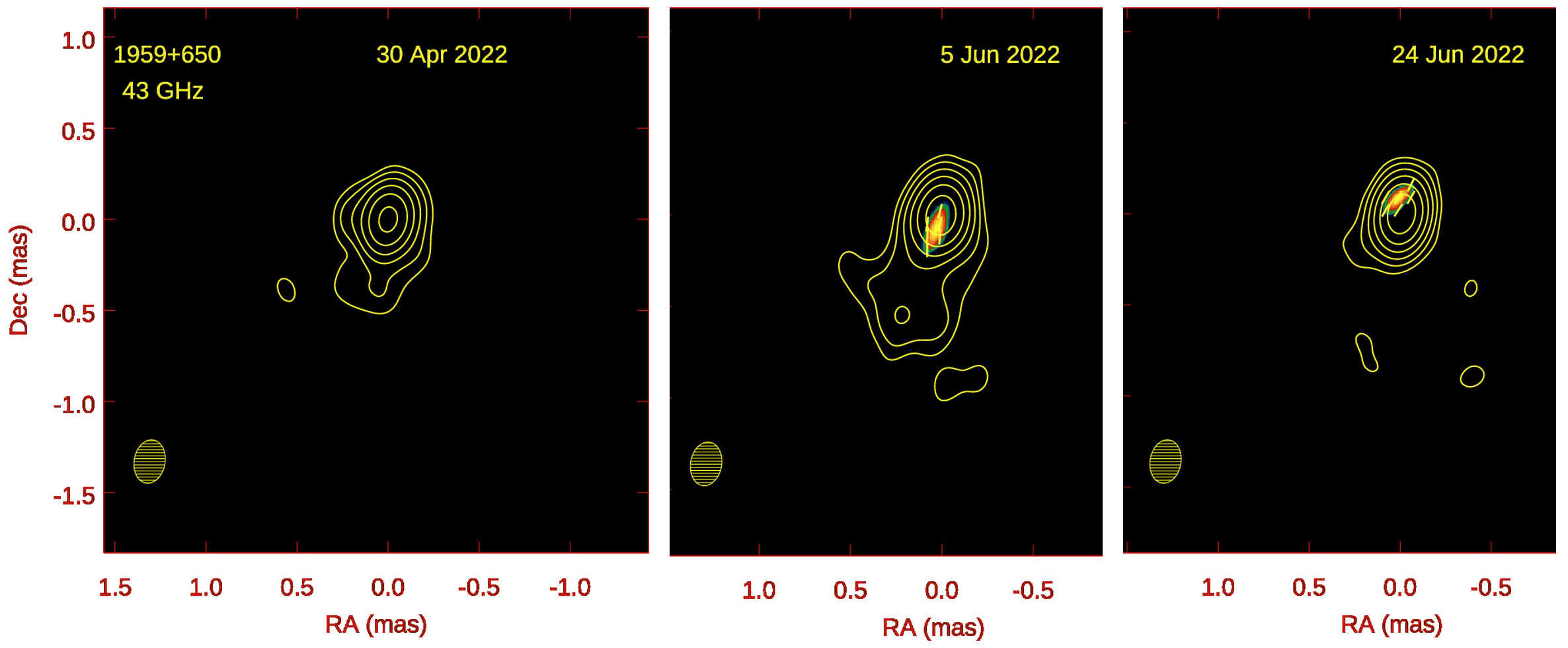}
\caption{VLBA images of \es\ at 7 mm at three epochs contemporaneous with the \ixpe\ observations. Contours: total
intensity in factors of 2, starting at 2\% of the peak intensity of (left to right) 0.16, 0.12, and 0.14 Jy per beam.
Color coding: relative linearly polarized intensity, with
maxima of (left to right) $<4$, 2.8, and 4.0 mJy per beam. Yellow
line segments: electric-vector position angle of polarization. Cross-hatched ellipse in the lower left of
each panel: FWHM of the elliptical restoring
beam corresponding to the angular
resolution along different directions. All positions are relative to the location 
of maximum intensity.}
\label{vlba.fig}
\end{figure*}

Images of \es\ obtained at 7\,mm (43\,GHz) from VLBA data feature a $\sim 1$ milliarcsecond jet extending to the southeast of a compact ``core'' along a position angle of $\phi \sim 150^\circ$ \citep{2010ApJ...723.1150P,weaver2022}. Figure~\ref{vlba.fig} presents VLBA images at three epochs near the \ixpe\ pointings obtained as part of the BEAM-ME monitoring project\footnote{www.bu.blazars/BEAM-ME.html}. The data were obtained and analyzed using standard procedures, as described by \citet{jorstad2017}. We measure the linear polarization of the core to be $P_{\rm core}=2.3\%\pm0.2\%$ along $\psi_{\rm core}=173\pm10\degr$ on 2022 June 5, and $P_{\rm core}=2.8\%\pm0.3\%$ along $\psi_{\rm core}=152\pm10\degr$ on 2022 June 24. We determine only an upper limit to the degree of polarization of $P_{\rm core}\leq2.4\%$ on 2022 April 30. When the resolved 43\,GHz emission is included, the degree of polarization decreases to $P=1.5\%\pm0.5\%$ and $P=1.7\%\pm0.9\%$  on June 5 and 24, respectively. 

The IRAM observations were obtained during the first \ixpe\ observation on 2022 May 5 (MJD~59705.0138) and during the second observation on 2022 June 12 (MJD~59742.9538) at 86.24 and 228.93 GHz, under the Polarimetric Monitoring of AGN at Millimeter Wavelengths (POLAMI) Large Program\footnote{\url{http://polami.iaa.es/}} \citep{Agudo2018, Agudo2018-II, Thum2018}. For the first observation, we obtained a 99\% upper limit of $\Pi_\mathrm{R}<8\%$ for the polarization degree at 228~GHz. During the second \ixpe\ measurement, the IRAM observations yielded upper limits of $\Pi_\mathrm{R}<1.6\%$ and $<9.6\%$ at 86\,GHz and 228\,GHz, respectively.

{SMA observations were obtained between the two IXPE exposures on 2022 June 1 (MJD~59731.60152) and 2022 June 6 (MJD~59736.71253) at 225.3~GHz within the SMAPOL (SMA Monitoring of AGNs with POLarization) program. Data were taken with the SMA polarimeter \citep{Marrone2008} in full polarization mode using the SWARM correlator \citep{Primiani2016}. MWC~349\,A, and Callisto were used as the total flux calibrators and 3C~286 as the polarization calibrator. We find a low degree of polarization $\Pi_\mathrm{R}=1.92\pm0.58\%$ along $\psi_\mathrm{R}=129\degr\pm8\degr$  for the first observation, and $\Pi_\mathrm{R}=2.53\pm0.5\%$ along $\psi_\mathrm{R}=156\degr\pm5\degr$ for the second one.}

IR photometric and polarimetric data in H (1.6\,$\mu$m) and K (2.2\,$\mu$m) bands during the \two\ \ixpe\ exposures were obtained using the MIMIR camera\footnote{https://people.bu.edu/clemens/mimir/index.html} at the Perkins Telescope (PTO, Flagstaff, AZ), operated by Boston University. The average and standard deviation of the host-corrected IR polarization degree was found to be $\Pi_\mathrm{H}=3.27\%\pm0.16\%$ along a polarization angle $\psi_\mathrm{H}=162\degr\pm25\degr$ and $\Pi_\mathrm{K}=4.73\%\pm0.75\%$ along $\psi_\mathrm{K}=151\degr\pm28\degr$. 

During the first \ixpe\ observation, we only have a single optical measurement from NOT, which yielded an optical polarization degree $\rm \Pi_{O}=4.49\pm0.17\%$ along $\rm \psi_O=159\pm1\degr$. During the second \ixpe\ observation, the median and standard deviation of the optical polarization degree was $\rm \Pi_O=5.4\%\pm1.1\%$ along $\rm \psi_O=103\degr\pm6\degr$. The IR and optical observations and data analysis, as well as the host-galaxy modeling, are described in Appendix~\ref{app:infrare/optical}.

\section{Implications for particle acceleration scenarios}
\label{sec:discussion}
The spectral energy distribution of the synchrotron emission from \es\  typically peaks in the X-ray band \citep[$\sim 0.1-100\,$keV;][]{2019A&A...632A..77C}. It belongs to the category of high-frequency-peaked BL~Lac-type blazars, a subclass that includes Mrk~501 and Mrk~421, which exhibit the highest X-ray fluxes \citep{1996ApJS..104..251P}. 
In a leptonic-dominated emission scenario, the IR to X-ray emission from these objects is attributed to synchrotron radiation produced by electrons (and possibly positrons) in the jet's magnetic field. 
Synchrotron radiation is inherently polarized, with a degree of polarization of $\Pi \sim 70\%$ for electrons in a homogeneous magnetic field. After correcting for relativistic aberration, the polarization angle ($\psi$) at optically thin wavelengths is perpendicular to the average magnetic field direction projected onto the plane of the sky.
In the case of BL~Lac-type blazars, observations typically show polarization degrees of $\Pi \lesssim 20\%$ \citep{Jorstad2005AJ....130.1418J, Smith2007ApJ...663..118, Hovatta2016, Blinov2021}.

There are two main reasons for the reduction
in observed polarization.
First, geometrical depolarization occurs when there is a random component in the magnetic field with varying orientations within the emitting region. This geometrical depolarization effect becomes more pronounced as the ratio of random to ordered magnetic field ($B_{r}/B_{0}$) increases \citep{1965ARA&A...3..297G}.
Second, temporal depolarization happens when the magnetic field direction changes over timescales shorter than integration time of the polarization measurement. The magnitude of this effect increases with the amplitude of unresolved changes in polarization angle \citep{DiGesu2022models, zhang2022}. This phenomenon was observed in \ixpe\ observations of Mrk~421 \citep{digesuRotation}.

The degree of polarization in the X-ray band can be compared to that measured at optical frequencies during the same epoch. This polarization ratio $\Pi_\mathrm{x}/\Pi_\mathrm{O}$ can be used as a diagnostic for potential differences in the magnetic field geometry and dynamics encountered by the populations of electrons responsible for the emission in each band. Previous \ixpe\ observations of Mrk~501 and Mrk~421 have shown higher X-ray polarization compared to optical and radio frequencies \citep{Liodakis2022-Mrk501,DiGesu2022-Mrk421}, which can be explained by an energy-stratified particle distribution that would naturally arise in shock acceleration scenarios \citep[e.g.,][]{tavecchio2018}. 

In 2022 \one, the ratio of X-ray to optical degree of polarization in \es\ was measured to be $\Pi_\mathrm{x}/\Pi_\mathrm{O} = 1.8 \pm 0.5$, lower than observed in Mrk~421 \citep[$\Pi_\mathrm{x}/\Pi_\mathrm{O} \sim 5$;][]{DiGesu2022-Mrk421} and comparable to Mrk~501 \citep[$\Pi_\mathrm{x}/\Pi_\mathrm{O} \sim 2.5$;][]{Liodakis2022-Mrk501}. 
As in Mrk~501, the X-ray and optical polarization angles of \es\ were aligned (within $30^\circ$) with the jet direction, inferred to be $\phi\sim 150^\circ$ from VLBA images \citep[][Figure~\ref{vlba.fig}]{2010ApJ...723.1150P,weaver2022}, indicating a magnetic field geometry consistent with a helical or toroidal configuration \citep{2012AJ....144..105H} or compression by a shock \citep{Hughes1985}. 

In contrast, during the second \ixpe\ exposure 
in 2022 \two, the average optical polarization was $\Pi_\mathrm{O}=5.4\% \pm 1.1\%$, while the X-ray polarization was $\Pi_\mathrm{x} < 5.1\%$ (Figure~\ref{fig:pol_obs}), suggesting $\Pi_\mathrm{x}/\Pi_\mathrm{o} \lesssim 1.0$. The optical polarization angle exhibited a rotation from $\psi_\mathrm{O} \gtrsim 170^\circ$ one day before \two\ to $103^\circ$ during the \ixpe\ exposure, returning to $\sim 160^\circ$ five days after (Figure~\ref{fig:pol_light}). Similar polarization angle excursions were observed in 2013-2016 \citep{Blinov2021}. IR observations also indicated a change in polarization angle during the second \ixpe\ exposure. The X-ray flux exhibited smooth variations with a peak-to-valley amplitude of approximately $50\%$ of the median X-ray flux (Figure~\ref{fig:pol_light}).  

Various {\rev single-zone} models can explain the observed depolarization as geometrical. In shock acceleration scenarios, energy-dependent geometrical depolarization arises as electrons with decreasing energies occupy a larger volume, leading to a higher random magnetic field component and increased geometrical depolarization toward lower frequencies \citep{Angelakis2016,tavecchio2018,Liodakis2022-Mrk501}. 
In magnetic reconnection scenarios, geometrical depolarization can occur due to the increasing range of magnetic field orientations as high-energy electrons move away from X-points where the highly-ordered magnetic field changes direction \citep{2009ApJ...699.1789T,2014ApJ...783L..21S}. 
However, a precise calculation of the X-ray versus optical polarization degree is not straightforward. 
The presence of a kink instability can create preferred locations for current sheets, and if the reconnecting fields are mainly toroidal \citep{zhang2022}, the emitted radiation should exhibit a polarization angle roughly aligned with the jet axis.

Multi-zone models with turbulent magnetic field cells predict greater geometrical depolarization at optical frequencies than at X-ray energies \citep{2014ApJ...780...87M}. However, in such models it is the {\it mean} X-ray polarization that should exceed the {\it mean} optical polarization by a factor $\gtrsim2$
\citep{marscher2022}. Temporal fluctuations of both, with a standard deviation equal to 0.5 times the mean, are expected, so the $\Pi_{\mathrm X}/\Pi_{\mathrm O}$ ratio should vary over time.
Turbulence-based models also predict polarization angle swings that are not associated with changes in flux or polarization degree \citep{2023arXiv230113316Z}.

\section{Conclusions}
\label{sec:conclusions}
In 2022 \one, \ixpe\ detected X-ray polarization from the BL Lac-type blazar \es. The degree of polarization in the 2-8\,keV band was measured to be $\Pi_\mathrm{x}=8.0 \pm 2.3$, with an electric-vector polarization position angle $\psi_\mathrm{x}=123^\circ \pm 8^\circ$. \citet{2023arXiv231101745B} do not report a significant detection of X-ray polarization from this data set, possibly due to their use of a model-independent analysis that is less sensitive than the spectro-polarimetric analysis presented here. 
Our study constrains the degree of polarization during a second \ixpe\ exposure in 2022 \two\ 
to $\Pi_\mathrm{x} < 5.1\%$. {\rev Subsequent observations of \es\ with \ixpe\ in October 2022 and August 2023 reveal increased X-ray polarization levels, reaching $\Pi_\mathrm{x}=12.5 \pm 0.7$  \citep{2023arXiv231101745B}, compared to the levels from the earlier campaign reported here.} 

A time-resolved analysis of the second data set shows significant polarization during one of four 50\,ks time bins, with polarization degree and angle compatible with the \one\ observation.  Considering the measurement uncertainties, the hypothesis of a constant $\Pi_\mathrm{x}$ over the two observations can be only marginally rejected at $2.3\sigma$ confidence level. 

During the 50-day period including the two \ixpe\ observations, optical polarization measurements showed a range of $2.3\% \leq \Pi_\mathrm{O} \leq  6.5\%$, with a median value of 3.9\%. 
When comparing the degree of polarization between the X-ray and optical bands, a ratio of $\Pi_\mathrm{x}/\Pi_\mathrm{O} = 1.8 \pm 0.5$ was observed during \one, while $\Pi_\mathrm{x}/\Pi_\mathrm{o} \lesssim 1.0$ was observed during \two. 

The amplitude of the galaxy-subtracted optical flux variability was 35\% (peak to valley) during the campaign, while the X-ray flux changed by a factor of $\sim3$. Although the optical and IR position angles of polarization varied significantly during the second \ixpe\ pointing, the degree of polarization in these bands remained low, at around 
$4\%$. 
This suggests the presence of multiple magnetic field directions within the optical and IR emission regions, which could be a result of turbulence or magnetic reconnection events.

Recent observations of nearby high-frequency-peaked BL~Lacs (Mrk~501 and Mrk~421, $z\sim0.03$) at X-ray and optical frequencies have shown a higher degree of polarization in the X-ray band. This has been interpreted as evidence for an energy-stratified electron population dominating the observed radiative output of the jet \citep{Liodakis2022-Mrk501,DiGesu2022-Mrk421}. Similar results have been obtained with analogs of those objects at higher redshift \citep[PG~1553+113, $z\sim0.4$; 1ES~0229+200, $z=0.140$;][]{ixpe1553,ixpe0229}. 

Our observations of \es\ represent the first time that X-ray polarization is detected from a high-frequency-peaked BL~Lac without clearly exhibiting frequency dependence. The observed ratio $\Pi_\mathrm{x}/\Pi_\mathrm{o} \lesssim 1.0$ in \es\ during the second \ixpe\ pointing is in apparent contradiction with a scenario where electrons emitting optical radiation occupy a larger volume with a more turbulent magnetic field compared to the higher energy electrons responsible for the X-ray emission. Nevertheless, the higher-amplitude flux variability in the X-ray band compared to that at optical frequencies suggests a more compact X-ray emitting region. 

Temporal depolarization induced by turbulence in the X-ray emitting region could potentially explain occasions when the observed X-ray and optical polarization becomes $\Pi_\mathrm{X}\approx \Pi_\mathrm{O}$. 
If changes in polarization angle occur on timescales shorter than the 50\,ks resolution of \ixpe\ at the observed flux levels, the degree of polarization in the X-ray band will be  suppressed.
Optical observations, with integration times of  $\lesssim 2$\,ks, are less susceptible to temporal depolarization caused by slowly varying magnetic field configurations.

The change in optical polarization angle observed during the second \ixpe\ pointing, in the absence of significant X-ray or optical flux variability, could be explained by multi-zone emission from turbulent plasma cells with typical dynamical timescales $<1$\,day. However, models based on turbulence alone do not predict alignment of the electric-vector polarization angle with the projected jet axis direction. During this campaign, we observed $\psi_\mathrm{x} \approx \psi_\mathrm{o}$ aligned within $30\degr$ of the projected jet direction, as measured with the VLBA. Models in which turbulent plasma crosses a shock have the potential to explain alignment with the jet as well as the time variability \citep[][]{2014ApJ...780...87M,tavecchio2018,tavecchio2021}.

A plausible alternative explanation for the lower degree of X-ray polarization measured in \es\ compared to Mrk~501 and Mrk~421 could be a minor contribution from the rising edge of the high-energy emission component of the spectral energy distribution to the observed 2-8\,keV X-ray flux. This second emission component consists of Compton-scattered photons that are expected to have a very low degree of polarization \citep{Krawczynski2012}. Thus, their presence in the \ixpe\ data would weaken the measured degree of X-ray polarization in that band. Despite the absence of spectral curvature in the \xrt\ and \xmm\ spectral models, uncertainties might allow for a slight contribution from Compton-scattered emission at the highest energies. However, it is worth noting that no significant (inverted) spectral break, which would indicate a second spectral X-ray component, was indicated in the model fitting.

Given the time variability of the flux and polarization, further \ixpe\ and multi-wavelength polarization and flux monitoring observations of \es\ are needed to probe the magnetic field geometry and the relationship between the emission regions at X-ray, optical, and IR wavelengths. During our campaign, the influence of geometrical and temporal
depolarization cannot be fully distinguished. 
At X-ray flux levels of $F_{2-8\,\mathrm{keV}} = 10^{-10}$\,\fluxcgs\ and $\Pi_\mathrm{x} \sim 8\%$, \ixpe\ has the capability to detect changes in the polarization degree within approximately 50\,ks. The sensitivity of \ixpe\ scales as $MDP_{99} \propto (\mathtt{exposure} \times F_{2-8\,\mathrm{keV}})^{-1/2}$, indicating  that variations in X-ray polarization down to levels of $\Pi_\mathrm{x} \lesssim 5\%$ 
could be resolved when the X-ray flux from \es\ is $F_{2-8\,\mathrm{keV}} \gtrsim 3 \times 10^{-10}$\,\fluxcgs, or approximately twice as bright as the flux levels during the reported campaign. This capability has been demonstrated recently with the detection by \ixpe\ of a $>360\degr$ X-ray polarization angle rotation in Mrk~421 \citep{digesuRotation}.

\facilities{Calar Alto, IRAM-30m, \ixpe, KANATA, NOT, Perkins, SNO, Skinakas, {SMA}, VLBA, \xrt, \xmm.}

\software{astropy \citep{2013A&A...558A..33A,2018AJ....156..123A,2022ApJ...935..167A},  
Sherpa \citep{2001SPIE.4477...76F},
\texttt{Xspec} \citep{1996ASPC..101...17A},
\texttt{ixpeobssim} \citep{2022SoftX..1901194B,2022ascl.soft10020B},
HEASoft {\rev \citep{2014ascl.soft08004N}},
SAS {\rev \citep{2004ASPC..314..759G}}.}

\clearpage\newpage

%\begin{acknowledgments}
The authors thank Alex Chen for useful discussions on the expectations for a polarized flux arising from magnetic reconnection events. The Imaging X-ray Polarimetry Explorer (\ixpe) is a joint US and Italian mission. The US contribution is supported by the National Aeronautics and Space Administration (NASA) and led and managed by its Marshall Space Flight Center (MSFC), with industry partner Ball Aerospace (contract NNM15AA18C).  
The Italian contribution is supported by the Italian Space Agency (Agenzia Spaziale Italiana, ASI) through contract ASI-OHBI-2017-12-I.0, agreements ASI-INAF-2017-12-H0 and ASI-INFN-2017.13-H0, and its Space Science Data Center (SSDC), and by the Istituto Nazionale di Astrofisica (INAF) and the Istituto Nazionale di Fisica Nucleare (INFN) in Italy.  This research used data products provided by the \ixpe\ Team (MSFC, SSDC, INAF, and INFN) and distributed with additional software tools by the High-Energy Astrophysics Science Archive Research Center (HEASARC), at NASA Goddard Space Flight Center (GSFC). We acknowledge financial support from ASI-INAF agreement n.\ 2022-14-HH.0. I. L. was supported by the NASA Postdoctoral Program at the Marshall Space Flight Center, administered by Oak Ridge Associated Universities under contract with NASA. The research at Boston University was supported in part by National Science Foundation grant AST-2108622 and by NASA Swift Guest Investigator grant 80NSSC22K0537, and Fermi Guest Investigator grants 80NSSC21K1917 and 80NSSC22K1571. 
This research has made use of data from the RoboPol program, a collaboration between Caltech, the University of Crete, IA-FORTH, IUCAA, the MPIfR, and the Nicolaus Copernicus University, which was conducted at Skinakas Observatory in Crete, Greece. The IAA-CSIC co-authors acknowledge financial support from the Spanish "Ministerio de Ciencia e Innovaci\'{o}n" (MCIN/AEI/ 10.13039/501100011033) through the Center of Excellence Severo Ochoa award for the Instituto de Astrof\'{i}sica de Andaluc\'{i}a-CSIC (CEX2021-001131-S), and through grants PID2019-107847RB-C44 and PID2022-139117NB-C44. Some of the data are based on observations collected at the Observatorio de Sierra Nevada, owned and operated by the Instituto de Astrof\'{i}sica de Andaluc\'{i}a (IAA-CSIC). 
Further data are based on observations collected at the Centro Astron\'{o}mico Hispano-Alem\'{a}n(CAHA), operated jointly by Junta de Andaluc\'{i}a and Consejo Superior de Investigaciones Cient\'{i}ficas (IAA-CSIC). The POLAMI observations were carried out at the IRAM 30m Telescope. IRAM is supported by INSU/CNRS (France), MPG (Germany) and IGN (Spain). 
The data in this study include observations made with the Nordic Optical Telescope, owned in collaboration by the University of Turku and Aarhus University, and operated jointly by Aarhus University, the University of Turku and the University of Oslo, representing Denmark, Finland and Norway, the University of Iceland and Stockholm University at the Observatorio del Roque de los Muchachos, La Palma, Spain, of the Instituto de Astrofisica de Canarias. The data presented here were obtained in part with ALFOSC, which is provided by the Instituto de Astrof\'{\i}sica de Andaluc\'{\i}a (IAA) under a joint agreement with the University of Copenhagen and NOT. E.\ L.\ was supported by Academy of Finland projects 317636 and 320045. This study was based in part on observations conducted using the Perkins Telescope Observatory (PTO) in Arizona, USA, which is owned and operated by Boston University. 
We acknowledge funding to support our NOT observations from the Finnish Centre for Astronomy with ESO (FINCA), University of Turku, Finland (Academy of Finland grant nr 306531). Part of the French contribution is supported by the Scientific Research National Center (CNRS) and the French Spatial Agency (CNES).  D.B., S.K., R.S., N. M., acknowledge support from the European Research Council (ERC) under the European Unions Horizon 2020 research and innovation programme under grant agreement No.~771282. CC acknowledges support by the European Research Council (ERC) under the HORIZON ERC Grants 2021 programme under grant agreement No. 101040021. 
The VLBA is an instrument of the National Radio Astronomy Observatory. The National Radio Astronomy Observatory is a facility of the National Science Foundation operated under cooperative agreement by Associated Universities, Inc. This work was supported by JST, the establishment of university fellowships towards the creation of science technology innovation, Grant Number JPMJFS2129. 
This work was supported by Japan Society for the Promotion of Science (JSPS) KAKENHI Grant Numbers JP21H01137. This work was also partially supported by the Optical and Near-Infrared Astronomy Inter-University Cooperation Program from the Ministry of Education, Culture, Sports, Science and Technology (MEXT) of Japan. We are grateful to the observation and operating members of the Kanata Telescope. {The Submillimeter Array is a joint project between the Smithsonian Astrophysical Observatory and the Academia Sinica Institute of Astronomy and Astrophysics and is funded by the Smithsonian Institution and the Academia Sinica. Maunakea, the location of the SMA, is a culturally important site for the indigenous Hawaiian people; we are privileged to study the cosmos from its summit.}

%\end{acknowledgments}

\appendix

\section{X-ray data reduction}\label{app:x}
\subsection{\ixpe}
\textit{Level 2} \ixpe\ event files containing Stokes parameters calculated on an event-by-event basis were downloaded from the public \texttt{HEASARC} data repository. Data were further processed using \texttt{ixpeobssim~v30.2.2} \citep{2022SoftX..1901194B,2022ascl.soft10020B}. Signal events were extracted using the \texttt{xpselect} tool from a $80''$-radius circle centered on the sky location of \es. Background photon counts were extracted from an annulus with inner radius of $140''$ and outer radius of $200''$ centered on the same location. This procedure was repeated for each of the three \ixpe\ detector units \citep[DU1--3,][]{weisskopf2022}. Spectra were background-dominated at energies above 6\,keV. Stokes $I$, $Q$, and $U$ spectra in the 2-6\,keV band were therefore extracted using \texttt{xpbin} without using event weights and binned to a minimum of 8 counts per bin. Spectropolarimetric analysis was performed using \texttt{Sherpa v4.14} \citep{sherpa_2022_7186379}, with custom spectro-polarimetry models equivalent to the \texttt{polconst} models found in \texttt{Xspec}. 
Normalization constants for detector units aFor \two, the constant associated with \xmm\ is 0.82 and that of \xrt\ is 1.35, while those for detector units DU2 and DU3 are 0.97 and 0.92, all normalized to \ixpe\ DU1. The wider spread in inter-calibration constants can be attributed to X-ray flux variability of \es\ during and after \two.

\subsection{\xrt}
\xrt\ observations were performed both in photon counting and windowed timing mode. Files were processed using \texttt{xrtpipeline} packaged in \texttt{HEASoft v6.30} using the latest \xrt\ calibration files. Signal events were extracted from a $60''$-radius circle centered on \es. Background photon counts were extracted from an annulus with inner radius of $150''$ and outer radius of $280''$. Spectra were binned to have at least 40 counts per energy bin and fit using 
\texttt{Xspec} \citep{1996ASPC..101...17A} to extract fluxes and spectral  information.

\subsection{\xmm}
\xmm\ observations were performed with the PN camera in timing mode, using the thick filter to mitigate pile-up effects. EPAT plots clearly indicated the presence of pile-up in MOS cameras, rendering the MOS data unsuitable for analysis. Spectra  were generated using version \texttt{20210317\_1624-19.1.0} of the XMM-Newton Science Analysis Software (SAS) and the latest calibration products. To derive the spectrum of the source, a 27-pixel size box was employed, centered on the source itself. For background extraction, a region of the same size was selected from a blank area on the EPIC-pn CCD camera. To ensure accurate analysis, the resulting spectrum was appropriately rebinned to achieve a minimum of 30 counts per bin. Moderate pile-up was observed in the PN data below 1,keV, prompting the restriction of spectral analysis to the 1.0-10\,keV energy range. After background subtraction, the net exposure in the PN camera is $1.39 \times 10^4$\,s. 

\section{Infrared/Optical observations and data reduction}\label{app:infrare/optical}

\subsection{Infrared observations}

The IR observations were obtained with the MIMIR instrument at the 1.8m Perkins telescope in Flagstaff, Arizona. A detailed description of the camera and data reduction can be found in \cite{Clemens2012}. One observation was the result of 6 dithering exposures of 10 s each at 16 positions of a half-wave plate, rotated in steps of 22.5\degr\ from 0\degr\ to 360\degr.
Therefore, to determine polarization parameters at a given epoch, we collected 96 measurements of the source. We carried out 15 and 10 observations in H and K bands, respectively, from MJD 59736.5 to MJD 59748.5. To perform photometry, we have co-added all 96 exposures at a given epoch to construct a deep image. These deep images were also used to model the host galaxy of 1959+650 in H and K bands. 

To make a model of the host galaxy, we used a photometric decomposition technique, which allows one to find an analytical model of an object and separate the light coming from its components. The distance to the object does not allow us to resolve it in great detail on our IR images, so in this work we adopted a rather simple model consisting of two components:
a Sersic \citep{Sersic1963, Sersic1968} function to describe the host galaxy, and a point source for the active nucleus. We built the point-spread function (PSF) by averaging images of bright, isolated, and unsaturated stars, and fit the resulting image with a Moffat function \citep{Moffat1969} to produce a smooth image. To run the decomposition, we used the {\small IMFIT} 
package \citep{Erwin2015}. Our estimates of the host galaxy structural parameters are as follows: ellipticity $e=0.31 \pm 0.05$, position angle $PA=72.9 \pm 13.1\degr$, Sersic index $n=1.55 \pm 0.5$, effective radius $r_e(K)=2.53 \pm 0.46'', r_e(H)=2.84 \pm 0.35''$, effective surface brightness $\mu_e(K)=16.98 \pm 0.39''$, and $\mu_e(H)=17.65 \pm 0.27''$ mag per square arcsecond. The rather large uncertainties, especially in the Sersic index value, are attributed to poor resolution of the object: the observed object effective radius is close to the PSF FWHM, which results in a degradation of the decomposition quality \citep{Trujillo2001}. We have employed the model parameters to estimate  contributions of the host galaxy of 1959+650 within an aperture of diameter $5''$ that are equal to $14.28\pm0.09$\,mag and $13.71\pm0.05$\,mag in H and K bands, respectively. 

\subsection{Optical Observations}

The optical R-band observations were performed at the Calar Alto observatory, KANATA telescope, the Nordic Optical Telescope (NOT) using the Alhambra Faint Object Spectrograph and Camera (ALFOSC), the Sierra Nevada observatory, and at the Skinakas observatory using the RoboPol polarimeter. The Calar Alto and SNO observations were obtained on 2022 June 6, 10, and 17 (MJD~59737.1, 59740.9, 59741.0, and 59747.8), and analyzed using standard photo-polarimetric procedures. The KANATA observations were obtained on 2022 May 3 (MJD~59702.0) using the Hiroshima Optical and Near-InfraRed camera \citep[HONIR,][]{Akitaya2014,Kawabata1999}, and analyzed using standard procedures.  The NOT observations were obtained on 2022 May 4 (MJD~59703.1) and 2022 June 7 (MJD~59737.1). The data were analyzed using unpolarized and polarized standard stars for calibration, following the standard photometric procedures included in the Tuorla Observatory pipeline \citep{Hovatta2016,Nilsson2018}. The RoboPol polarimeter is a novel 4-channel polarimeter that simultaneously measures $\Pi_O$ and $\psi_O$ with a single exposure \citep{Ramaprakash2019}. Details of the analysis pipeline and data reduction procedures can be found in \citet{Panopoulou2015} and \citet{Blinov2021}.

All of the optical observations were corrected for the depolarization effect of unpolarized host-galaxy contribution to the emission following \cite{Nilsson2007}. A more detailed description of the analysis procedures from the different observatories can be found in \cite{Liodakis2022-Mrk501,DiGesu2022-Mrk421,Middei2023}.

\bibliography{main}{}

\begin{thebibliography}{}
\expandafter\ifx\csname natexlab\endcsname\relax\def\natexlab#1{#1}\fi
\providecommand{\url}[1]{\href{#1}{#1}}
\providecommand{\dodoi}[1]{doi:~\href{http://doi.org/#1}{\nolinkurl{#1}}}
\providecommand{\doeprint}[1]{\href{http://ascl.net/#1}{\nolinkurl{http://ascl.net/#1}}}
\providecommand{\doarXiv}[1]{\href{https://arxiv.org/abs/#1}{\nolinkurl{https://arxiv.org/abs/#1}}}

\bibitem[{{Agudo} {et~al.}(2018{\natexlab{a}}){Agudo}, {Thum}, {Ramakrishnan}, {Molina}, {Casadio}, \& {G{\'o}mez}}]{Agudo2018-II}
{Agudo}, I., {Thum}, C., {Ramakrishnan}, V., {et~al.} 2018{\natexlab{a}}, \mnras, 473, 1850, \dodoi{10.1093/mnras/stx2437}

\bibitem[{{Agudo} {et~al.}(2018{\natexlab{b}}){Agudo}, {Thum}, {Molina}, {Casadio}, {Wiesemeyer}, {Morris}, {Paubert}, {G{\'o}mez}, \& {Kramer}}]{Agudo2018}
{Agudo}, I., {Thum}, C., {Molina}, S.~N., {et~al.} 2018{\natexlab{b}}, \mnras, 474, 1427, \dodoi{10.1093/mnras/stx2435}

\bibitem[{{Akitaya} {et~al.}(2014){Akitaya}, {Moritani}, {Ui}, {Urano}, {Ohashi}, {Kawabata}, {Nakashima}, {Sasada}, {Sakimoto}, {Harao}, {Miyamoto}, {Matsui}, {Itoh}, {Takaki}, {Ueno}, {Ohsugi}, {Nakaya}, {Yamashita}, \& {Yoshida}}]{Akitaya2014}
{Akitaya}, H., {Moritani}, Y., {Ui}, T., {et~al.} 2014, in Society of Photo-Optical Instrumentation Engineers (SPIE) Conference Series, Vol. 9147, Ground-based and Airborne Instrumentation for Astronomy V, ed. S.~K. {Ramsay}, I.~S. {McLean}, \& H.~{Takami}, 91474O, \dodoi{10.1117/12.2054577}

\bibitem[{{Aliu} {et~al.}(2013){Aliu}, {Archambault}, {Arlen}, {Aune}, {Beilicke}, {Benbow}, {Bird}, {B{\"o}ttcher}, {Bouvier}, {Bugaev}, {Byrum}, {Cesarini}, {Ciupik}, {Collins-Hughes}, {Connolly}, {Cui}, {Dickherber}, {Duke}, {Dumm}, {Errand o}, {Falcone}, {Federici}, {Feng}, {Finley}, {Finnegan}, {Fortson}, {Furniss}, {Galante}, {Gall}, {Gillanders}, {Griffin}, {Grube}, {Gyuk}, {Hanna}, {Holder}, {Hughes}, {Humensky}, {Kaaret}, {Kertzman}, {Khassen}, {Kieda}, {Krawczynski}, {Krennrich}, {Lang}, {Madhavan}, {Maier}, {Majumdar}, {McArthur}, {McCann}, {Moriarty}, {Mukherjee}, {Nelson}, {O'Faol{\'a}in de Bhr{\'o}ithe}, {Ong}, {Orr}, {Otte}, {Park}, {Perkins}, {Pichel}, {Pohl}, {Popkow}, {Prokoph}, {Quinn}, {Ragan}, {Reyes}, {Reynolds}, {Roache}, {Saxon}, {Schroedter}, {Sembroski}, {Skole}, {Smith}, {Staszak}, {Telezhinsky}, {Theiling}, {Tyler}, {Varlotta}, {Vassiliev}, {Wakely}, {Weekes}, {Weinstein}, {Welsing}, {Williams}, \& {Zitzer}}]{2013ApJ...775....3A}
{Aliu}, E., {Archambault}, S., {Arlen}, T., {et~al.} 2013, \apj, 775, 3, \dodoi{10.1088/0004-637X/775/1/3}

\bibitem[{{Aliu} {et~al.}(2014){Aliu}, {Archambault}, {Arlen}, {Aune}, {Barnacka}, {Beilicke}, {Benbow}, {Berger}, {Bird}, {Bouvier}, {Buckley}, {Bugaev}, {Cerruti}, {Chen}, {Ciupik}, {Collins-Hughes}, {Connolly}, {Cui}, {Dumm}, {Eisch}, {Falcone}, {Federici}, {Feng}, {Finley}, {Fleischhack}, {Fortin}, {Fortson}, {Furniss}, {Galante}, {Gillanders}, {Griffin}, {Griffiths}, {Grube}, {Gyuk}, {H{\r{a}}kansson}, {Hanna}, {Holder}, {Hughes}, {Hughes}, {Humensky}, {Johnson}, {Kaaret}, {Kar}, {Kertzman}, {Khassen}, {Kieda}, {Krawczynski}, {Krennrich}, {Lang}, {Madhavan}, {Majumdar}, {McArthur}, {McCann}, {Meagher}, {Millis}, {Moriarty}, {Mukherjee}, {Nelson}, {Nieto}, {O'Faol{\'a}in de Bhr{\'o}ithe}, {Ong}, {Otte}, {Park}, {Perkins}, {Pohl}, {Popkow}, {Prokoph}, {Quinn}, {Ragan}, {Rajotte}, {Reyes}, {Reynolds}, {Richards}, {Roache}, {Sadun}, {Santander}, {Sembroski}, {Shahinyan}, {Sheidaei}, {Smith}, {Staszak}, {Telezhinsky}, {Theiling}, {Tyler}, {Varlotta}, {Vassiliev}, {Vincent}, {Wakely}, {Weekes}, {Weinstein},
  {Welsing}, {Wilhelm}, {Williams}, {Zitzer}, {VERITAS Collaboration}, {B{\"o}ttcher}, \& {Fumagalli}}]{2014ApJ...797...89A}
---. 2014, \apj, 797, 89, \dodoi{10.1088/0004-637X/797/2/89}

\bibitem[{{Angel} \& {Stockman}(1980)}]{1980ARA&A..18..321A}
{Angel}, J.~R.~P., \& {Stockman}, H.~S. 1980, \araa, 18, 321, \dodoi{10.1146/annurev.aa.18.090180.001541}

\bibitem[{{Angelakis} {et~al.}(2016){Angelakis}, {Hovatta}, {Blinov}, {Pavlidou}, {Kiehlmann}, {Myserlis}, {B{\"o}ttcher}, {Mao}, {Panopoulou}, {Liodakis}, {King}, {Balokovi{\'c}}, {Kus}, {Kylafis}, {Mahabal}, {Marecki}, {Paleologou}, {Papadakis}, {Papamastorakis}, {Pazderski}, {Pearson}, {Prabhudesai}, {Ramaprakash}, {Readhead}, {Reig}, {Tassis}, {Urry}, \& {Zensus}}]{Angelakis2016}
{Angelakis}, E., {Hovatta}, T., {Blinov}, D., {et~al.} 2016, \mnras, 463, 3365, \dodoi{10.1093/mnras/stw2217}

\bibitem[{{Arnaud}(1996)}]{1996ASPC..101...17A}
{Arnaud}, K.~A. 1996, in Astronomical Society of the Pacific Conference Series, Vol. 101, Astronomical Data Analysis Software and Systems V, ed. G.~H. {Jacoby} \& J.~{Barnes}, 17

\bibitem[{{Astropy Collaboration} {et~al.}(2013){Astropy Collaboration}, {Robitaille}, {Tollerud}, {Greenfield}, {Droettboom}, {Bray}, {Aldcroft}, {Davis}, {Ginsburg}, {Price-Whelan}, {Kerzendorf}, {Conley}, {Crighton}, {Barbary}, {Muna}, {Ferguson}, {Grollier}, {Parikh}, {Nair}, {Unther}, {Deil}, {Woillez}, {Conseil}, {Kramer}, {Turner}, {Singer}, {Fox}, {Weaver}, {Zabalza}, {Edwards}, {Azalee Bostroem}, {Burke}, {Casey}, {Crawford}, {Dencheva}, {Ely}, {Jenness}, {Labrie}, {Lim}, {Pierfederici}, {Pontzen}, {Ptak}, {Refsdal}, {Servillat}, \& {Streicher}}]{2013A&A...558A..33A}
{Astropy Collaboration}, {Robitaille}, T.~P., {Tollerud}, E.~J., {et~al.} 2013, \aap, 558, A33, \dodoi{10.1051/0004-6361/201322068}

\bibitem[{{Astropy Collaboration} {et~al.}(2018){Astropy Collaboration}, {Price-Whelan}, {Sip{\H{o}}cz}, {G{\"u}nther}, {Lim}, {Crawford}, {Conseil}, {Shupe}, {Craig}, {Dencheva}, {Ginsburg}, {VanderPlas}, {Bradley}, {P{\'e}rez-Su{\'a}rez}, {de Val-Borro}, {Aldcroft}, {Cruz}, {Robitaille}, {Tollerud}, {Ardelean}, {Babej}, {Bach}, {Bachetti}, {Bakanov}, {Bamford}, {Barentsen}, {Barmby}, {Baumbach}, {Berry}, {Biscani}, {Boquien}, {Bostroem}, {Bouma}, {Brammer}, {Bray}, {Breytenbach}, {Buddelmeijer}, {Burke}, {Calderone}, {Cano Rodr{\'\i}guez}, {Cara}, {Cardoso}, {Cheedella}, {Copin}, {Corrales}, {Crichton}, {D'Avella}, {Deil}, {Depagne}, {Dietrich}, {Donath}, {Droettboom}, {Earl}, {Erben}, {Fabbro}, {Ferreira}, {Finethy}, {Fox}, {Garrison}, {Gibbons}, {Goldstein}, {Gommers}, {Greco}, {Greenfield}, {Groener}, {Grollier}, {Hagen}, {Hirst}, {Homeier}, {Horton}, {Hosseinzadeh}, {Hu}, {Hunkeler}, {Ivezi{\'c}}, {Jain}, {Jenness}, {Kanarek}, {Kendrew}, {Kern}, {Kerzendorf}, {Khvalko}, {King}, {Kirkby}, {Kulkarni},
  {Kumar}, {Lee}, {Lenz}, {Littlefair}, {Ma}, {Macleod}, {Mastropietro}, {McCully}, {Montagnac}, {Morris}, {Mueller}, {Mumford}, {Muna}, {Murphy}, {Nelson}, {Nguyen}, {Ninan}, {N{\"o}the}, {Ogaz}, {Oh}, {Parejko}, {Parley}, {Pascual}, {Patil}, {Patil}, {Plunkett}, {Prochaska}, {Rastogi}, {Reddy Janga}, {Sabater}, {Sakurikar}, {Seifert}, {Sherbert}, {Sherwood-Taylor}, {Shih}, {Sick}, {Silbiger}, {Singanamalla}, {Singer}, {Sladen}, {Sooley}, {Sornarajah}, {Streicher}, {Teuben}, {Thomas}, {Tremblay}, {Turner}, {Terr{\'o}n}, {van Kerkwijk}, {de la Vega}, {Watkins}, {Weaver}, {Whitmore}, {Woillez}, {Zabalza}, \& {Astropy Contributors}}]{2018AJ....156..123A}
{Astropy Collaboration}, {Price-Whelan}, A.~M., {Sip{\H{o}}cz}, B.~M., {et~al.} 2018, \aj, 156, 123, \dodoi{10.3847/1538-3881/aabc4f}

\bibitem[{{Astropy Collaboration} {et~al.}(2022){Astropy Collaboration}, {Price-Whelan}, {Lim}, {Earl}, {Starkman}, {Bradley}, {Shupe}, {Patil}, {Corrales}, {Brasseur}, {N{\"o}the}, {Donath}, {Tollerud}, {Morris}, {Ginsburg}, {Vaher}, {Weaver}, {Tocknell}, {Jamieson}, {van Kerkwijk}, {Robitaille}, {Merry}, {Bachetti}, {G{\"u}nther}, {Aldcroft}, {Alvarado-Montes}, {Archibald}, {B{\'o}di}, {Bapat}, {Barentsen}, {Baz{\'a}n}, {Biswas}, {Boquien}, {Burke}, {Cara}, {Cara}, {Conroy}, {Conseil}, {Craig}, {Cross}, {Cruz}, {D'Eugenio}, {Dencheva}, {Devillepoix}, {Dietrich}, {Eigenbrot}, {Erben}, {Ferreira}, {Foreman-Mackey}, {Fox}, {Freij}, {Garg}, {Geda}, {Glattly}, {Gondhalekar}, {Gordon}, {Grant}, {Greenfield}, {Groener}, {Guest}, {Gurovich}, {Handberg}, {Hart}, {Hatfield-Dodds}, {Homeier}, {Hosseinzadeh}, {Jenness}, {Jones}, {Joseph}, {Kalmbach}, {Karamehmetoglu}, {Ka{\l}uszy{\'n}ski}, {Kelley}, {Kern}, {Kerzendorf}, {Koch}, {Kulumani}, {Lee}, {Ly}, {Ma}, {MacBride}, {Maljaars}, {Muna}, {Murphy}, {Norman},
  {O'Steen}, {Oman}, {Pacifici}, {Pascual}, {Pascual-Granado}, {Patil}, {Perren}, {Pickering}, {Rastogi}, {Roulston}, {Ryan}, {Rykoff}, {Sabater}, {Sakurikar}, {Salgado}, {Sanghi}, {Saunders}, {Savchenko}, {Schwardt}, {Seifert-Eckert}, {Shih}, {Jain}, {Shukla}, {Sick}, {Simpson}, {Singanamalla}, {Singer}, {Singhal}, {Sinha}, {Sip{\H{o}}cz}, {Spitler}, {Stansby}, {Streicher}, {{\v{S}}umak}, {Swinbank}, {Taranu}, {Tewary}, {Tremblay}, {de Val-Borro}, {Van Kooten}, {Vasovi{\'c}}, {Verma}, {de Miranda Cardoso}, {Williams}, {Wilson}, {Winkel}, {Wood-Vasey}, {Xue}, {Yoachim}, {Zhang}, {Zonca}, \& {Astropy Project Contributors}}]{2022ApJ...935..167A}
{Astropy Collaboration}, {Price-Whelan}, A.~M., {Lim}, P.~L., {et~al.} 2022, \apj, 935, 167, \dodoi{10.3847/1538-4357/ac7c74}

\bibitem[{{Baldini} {et~al.}(2022{\natexlab{a}}){Baldini}, {Bucciantini}, {Lalla}, {Ehlert}, {Manfreda}, {Negro}, {Omodei}, {Pesce-Rollins}, {Sgr{\`o}}, \& {Silvestri}}]{2022SoftX..1901194B}
{Baldini}, L., {Bucciantini}, N., {Lalla}, N.~D., {et~al.} 2022{\natexlab{a}}, SoftwareX, 19, 101194, \dodoi{10.1016/j.softx.2022.101194}

\bibitem[{{Baldini} {et~al.}(2022{\natexlab{b}}){Baldini}, {Bucciantini}, {Lalla}, {Ehlert}, {Manfreda}, {Negro}, {Omodei}, {Pesce-Rollins}, {Sgr{\`o}}, \& {Silvestri}}]{2022ascl.soft10020B}
---. 2022{\natexlab{b}}, {ixpeobssim: Imaging X-ray Polarimetry Explorer simulator and analyzer}, Astrophysics Source Code Library, record ascl:2210.020.
\newblock \doeprint{2210.020}

\bibitem[{{Begelman} {et~al.}(1984){Begelman}, {Blandford}, \& {Rees}}]{1984RvMP...56..255B}
{Begelman}, M.~C., {Blandford}, R.~D., \& {Rees}, M.~J. 1984, Reviews of Modern Physics, 56, 255, \dodoi{10.1103/RevModPhys.56.255}

\bibitem[{{Bharathan} {et~al.}(2023){Bharathan}, {Stalin}, {Chatterjee}, {Sahayanathan}, {Pal}, {Mathew}, \& {Agrawal}}]{2023arXiv231101745B}
{Bharathan}, A.~M., {Stalin}, C.~S., {Chatterjee}, R., {et~al.} 2023, arXiv e-prints, arXiv:2311.01745, \dodoi{10.48550/arXiv.2311.01745}

\bibitem[{{Blandford} {et~al.}(2019){Blandford}, {Meier}, \& {Readhead}}]{2019ARA&A..57..467B}
{Blandford}, R., {Meier}, D., \& {Readhead}, A. 2019, \araa, 57, 467, \dodoi{10.1146/annurev-astro-081817-051948}

\bibitem[{{Blandford} \& {Ostriker}(1978)}]{1978ApJ...221L..29B}
{Blandford}, R.~D., \& {Ostriker}, J.~P. 1978, \apjl, 221, L29, \dodoi{10.1086/182658}

\bibitem[{{Blandford} \& {Znajek}(1977)}]{Blandford1977}
{Blandford}, R.~D., \& {Znajek}, R.~L. 1977, \mnras, 179, 433, \dodoi{10.1093/mnras/179.3.433}

\bibitem[{{Blinov} {et~al.}(2021){Blinov}, {Kiehlmann}, {Pavlidou}, {Panopoulou}, {Skalidis}, {Angelakis}, {Casadio}, {Einoder}, {Hovatta}, {Kokolakis}, {Kougentakis}, {Kus}, {Kylafis}, {Kyritsis}, {Lalakos}, {Liodakis}, {Maharana}, {Makrydopoulou}, {Mandarakas}, {Maragkakis}, {Myserlis}, {Papadakis}, {Paterakis}, {Pearson}, {Ramaprakash}, {Readhead}, {Reig}, {S{\l}owikowska}, {Tassis}, {Xexakis}, {{\.Z}ejmo}, \& {Zensus}}]{Blinov2021}
{Blinov}, D., {Kiehlmann}, S., {Pavlidou}, V., {et~al.} 2021, \mnras, 501, 3715, \dodoi{10.1093/mnras/staa3777}

\bibitem[{{Bodo} {et~al.}(2021){Bodo}, {Tavecchio}, \& {Sironi}}]{Bodo2021}
{Bodo}, G., {Tavecchio}, F., \& {Sironi}, L. 2021, \mnras, 501, 2836, \dodoi{10.1093/mnras/staa3620}

\bibitem[{Burke {et~al.}(2022)Burke, Laurino, G\"unther, Marie-Terrell, Siemiginowska, Budynkiewicz, Cheer, Aldcroft, Deil, Sip{\H o}cz, Buchner, Donath, Laginja, Leinweber, nplee, Todd, wmclaugh, \& dtnguyen2}]{sherpa_2022_7186379}
Burke, D., Laurino, O., G\"unther, H.~M., {et~al.} 2022, sherpa/sherpa: Sherpa 4.15.0, 4.15.0,  Zenodo, \dodoi{10.5281/zenodo.7186379}

\bibitem[{{Burrows} {et~al.}(2005){Burrows}, {Hill}, {Nousek}, {Kennea}, {Wells}, {Osborne}, {Abbey}, {Beardmore}, {Mukerjee}, {Short}, {Chincarini}, {Campana}, {Citterio}, {Moretti}, {Pagani}, {Tagliaferri}, {Giommi}, {Capalbi}, {Tamburelli}, {Angelini}, {Cusumano}, {Br{\"a}uninger}, {Burkert}, \& {Hartner}}]{2005SSRv..120..165B}
{Burrows}, D.~N., {Hill}, J.~E., {Nousek}, J.~A., {et~al.} 2005, \ssr, 120, 165, \dodoi{10.1007/s11214-005-5097-2}

\bibitem[{{Chang} {et~al.}(2019){Chang}, {Arsioli}, {Giommi}, {Padovani}, \& {Brandt}}]{2019A&A...632A..77C}
{Chang}, Y.~L., {Arsioli}, B., {Giommi}, P., {Padovani}, P., \& {Brandt}, C.~H. 2019, \aap, 632, A77, \dodoi{10.1051/0004-6361/201834526}

\bibitem[{{Clemens} {et~al.}(2012){Clemens}, {Pinnick}, \& {Pavel}}]{Clemens2012}
{Clemens}, D.~P., {Pinnick}, A.~F., \& {Pavel}, M.~D. 2012, \apjs, 200, 20, \dodoi{10.1088/0067-0049/200/2/20}

\bibitem[{{de Vaucouleurs} {et~al.}(1991){de Vaucouleurs}, {de Vaucouleurs}, {Corwin}, {Buta}, {Paturel}, \& {Fouque}}]{1991rc3..book.....D}
{de Vaucouleurs}, G., {de Vaucouleurs}, A., {Corwin}, Herold~G., J., {et~al.} 1991, {Third Reference Catalogue of Bright Galaxies}

\bibitem[{{Dermer} {et~al.}(1996){Dermer}, {Miller}, \& {Li}}]{1996ApJ...456..106D}
{Dermer}, C.~D., {Miller}, J.~A., \& {Li}, H. 1996, \apj, 456, 106, \dodoi{10.1086/176631}

\bibitem[{{Di Gesu} {et~al.}(2022{\natexlab{a}}){Di Gesu}, {Tavecchio}, {Donnarumma}, {Marscher}, {Pesce-Rollins}, \& {Landoni}}]{DiGesu2022models}
{Di Gesu}, L., {Tavecchio}, F., {Donnarumma}, I., {et~al.} 2022{\natexlab{a}}, \aap, 662, A83, \dodoi{10.1051/0004-6361/202243168}

\bibitem[{{Di Gesu} {et~al.}(2022{\natexlab{b}}){Di Gesu}, {Donnarumma}, {Tavecchio}, {Agudo}, {Barnounin}, {Cibrario}, {Di Lalla}, {Di Marco}, {Escudero}, {Errando}, {Jorstad}, {Kim}, {Kouch}, {Liodakis}, {Lindfors}, {Madejski}, {Marshall}, {Marscher}, {Middei}, {Muleri}, {Myserlis}, {Negro}, {Omodei}, {Pacciani}, {Paggi}, {Perri}, {Puccetti}, {Antonelli}, {Bachetti}, {Baldini}, {Baumgartner}, {Bellazzini}, {Bianchi}, {Bongiorno}, {Bonino}, {Brez}, {Bucciantini}, {Capitanio}, {Castellano}, {Cavazzuti}, {Ciprini}, {Costa}, {De Rosa}, {Del Monte}, {Doroshenko}, {Dov{\v{c}}iak}, {Ehlert}, {Enoto}, {Evangelista}, {Fabiani}, {Ferrazzoli}, {Garcia}, {Gunji}, {Hayashida}, {Heyl}, {Iwakiri}, {Karas}, {Kitaguchi}, {Kolodziejczak}, {Krawczynski}, {La Monaca}, {Latronico}, {Maldera}, {Manfreda}, {Marin}, {Marinucci}, {Massaro}, {Matt}, {Mitsuishi}, {Mizuno}, {Ng}, {O'Dell}, {Oppedisano}, {Papitto}, {Pavlov}, {Peirson}, {Pesce-Rollins}, {Petrucci}, {Pilia}, {Possenti}, {Poutanen}, {Ramsey}, {Rankin}, {Ratheesh},
  {Romani}, {Sgr{\`o}}, {Slane}, {Soffitta}, {Spandre}, {Tamagawa}, {Taverna}, {Tawara}, {Tennant}, {Thomas}, {Tombesi}, {Trois}, {Tsygankov}, {Turolla}, {Vink}, {Weisskopf}, {Wu}, {Xie}, \& {Zane}}]{DiGesu2022-Mrk421}
{Di Gesu}, L., {Donnarumma}, I., {Tavecchio}, F., {et~al.} 2022{\natexlab{b}}, \apjl, 938, L7, \dodoi{10.3847/2041-8213/ac913a}

\bibitem[{{Di Gesu} {et~al.}(2023){Di Gesu}, {Marshall}, {Ehlert}, {Kim}, {Donnarumma}, {Tavecchio}, {Liodakis}, {Kiehlmann}, {Agudo}, {Jorstad}, {Muleri}, {Marscher}, {Puccetti}, {Middei}, {Perri}, {Pacciani}, {Negro}, {Romani}, {Di Marco}, {Blinov}, {Bourbah}, {Kontopodis}, {Mandarakas}, {Romanopoulos}, {Skalidis}, {Vervelaki}, {Casadio}, {Escudero}, {Myserlis}, {Gurwell}, {Rao}, {Keating}, {Kouch}, {Lindfors}, {Aceituno}, {Bernardos}, {Bonnoli}, {Casanova}, {Garc{\'\i}a-Comas}, {Ag{\'\i}s-Gonz{\'a}lez}, {Husillos}, {Marchini}, {Sota}, {Imazawa}, {Sasada}, {Fukazawa}, {Kawabata}, {Uemura}, {Mizuno}, {Nakaoka}, {Akitaya}, {Savchenko}, {Vasilyev}, {G{\'o}mez}, {Antonelli}, {Barnouin}, {Bonino}, {Cavazzuti}, {Costamante}, {Chen}, {Cibrario}, {De Rosa}, {Di Pierro}, {Errando}, {Kaaret}, {Karas}, {Krawczynski}, {Lisalda}, {Madejski}, {Malacaria}, {Marin}, {Marinucci}, {Massaro}, {Matt}, {Mitsuishi}, {O'Dell}, {Paggi}, {Peirson}, {Petrucci}, {Ramsey}, {Tennant}, {Wu}, {Bachetti}, {Baldini}, {Baumgartner},
  {Bellazzini}, {Bianchi}, {Bongiorno}, {Brez}, {Bucciantini}, {Capitanio}, {Castellano}, {Ciprini}, {Costa}, {Del Monte}, {Di Lalla}, {Doroshenko}, {Dov{\v{c}}iak}, {Enoto}, {Evangelista}, {Fabiani}, {Ferrazzoli}, {Garcia}, {Gunji}, {Hayashida}, {Heyl}, {Iwakiri}, {Kislat}, {Kitaguchi}, {Kolodziejczak}, {La Monaca}, {Latronico}, {Maldera}, {Manfreda}, {Ng}, {Omodei}, {Oppedisano}, {Papitto}, {Pavlov}, {Pesce-Rollins}, {Pilia}, {Possenti}, {Poutanen}, {Rankin}, {Ratheesh}, {Roberts}, {Sgr{\`o}}, {Slane}, {Soffitta}, {Spandre}, {Swartz}, {Tamagawa}, {Taverna}, {Tawara}, {Thomas}, {Tombesi}, {Trois}, {Tsygankov}, {Turolla}, {Vink}, {Weisskopf}, {Xie}, \& {Zane}}]{digesuRotation}
{Di Gesu}, L., {Marshall}, H.~L., {Ehlert}, S.~R., {et~al.} 2023, Nature Astronomy, 7, 1245, \dodoi{10.1038/s41550-023-02032-7}

\bibitem[{{Ehlert} {et~al.}(2023){Ehlert}, {Liodakis}, {Middei}, {Marscher}, {Tavecchio}, {Agudo}, {Kouch}, {Lindfors}, {Nilsson}, {Myserlis}, {Gurwell}, {Rao}, {Aceituno}, {Bonnoli}, {Casanova}, {Ag{\'\i}s-Gonz{\'a}lez}, {Escudero}, {Husillos}, {Otero Santos}, {Sota}, {Angelakis}, {Kraus}, {Keating}, {Antonelli}, {Bachetti}, {Baldini}, {Baumgartner}, {Bellazzini}, {Bianchi}, {Bongiorno}, {Bonino}, {Brez}, {Bucciantini}, {Capitanio}, {Castellano}, {Cavazzuti}, {Chen}, {Ciprini}, {Costa}, {De Rosa}, {Del Monte}, {Di Gesu}, {Di Lalla}, {Di Marco}, {Donnarumma}, {Doroshenko}, {Dov{\v{c}}iak}, {Enoto}, {Evangelista}, {Fabiani}, {Ferrazzoli}, {Garcia}, {Gunji}, {Hayashida}, {Heyl}, {Iwakiri}, {Jorstad}, {Kaaret}, {Karas}, {Kislat}, {Kitaguchi}, {Kolodziejczak}, {Krawczynski}, {La Monaca}, {Latronico}, {Maldera}, {Manfreda}, {Marin}, {Marinucci}, {Marshall}, {Massaro}, {Matt}, {Mitsuishi}, {Mizuno}, {Muleri}, {Negro}, {Ng}, {O'Dell}, {Omodei}, {Oppedisano}, {Papitto}, {Pavlov}, {Peirson}, {Perri}, {Pesce-Rollins},
  {Petrucci}, {Pilia}, {Possenti}, {Poutanen}, {Puccetti}, {Ramsey}, {Rankin}, {Ratheesh}, {Roberts}, {Romani}, {Sgr{\'o}}, {Slane}, {Soffitta}, {Spandre}, {Swartz}, {Tamagawa}, {Taverna}, {Tawara}, {Tennant}, {Thomas}, {Tombesi}, {Trois}, {Tsygankov}, {Turolla}, {Vink}, {Weisskopf}, {Wu}, {Xie}, \& {Zane}}]{ixpe0229}
{Ehlert}, S.~R., {Liodakis}, I., {Middei}, R., {et~al.} 2023, \apj, 959, 61, \dodoi{10.3847/1538-4357/ad05c4}

\bibitem[{{Erwin}(2015)}]{Erwin2015}
{Erwin}, P. 2015, \apj, 799, 226, \dodoi{10.1088/0004-637X/799/2/226}

\bibitem[{{Falcke} {et~al.}(2004){Falcke}, {K{\"o}rding}, \& {Markoff}}]{2004A&A...414..895F}
{Falcke}, H., {K{\"o}rding}, E., \& {Markoff}, S. 2004, \aap, 414, 895, \dodoi{10.1051/0004-6361:20031683}

\bibitem[{{Freeman} {et~al.}(2001){Freeman}, {Doe}, \& {Siemiginowska}}]{2001SPIE.4477...76F}
{Freeman}, P., {Doe}, S., \& {Siemiginowska}, A. 2001, in Society of Photo-Optical Instrumentation Engineers (SPIE) Conference Series, Vol. 4477, Astronomical Data Analysis, ed. J.-L. {Starck} \& F.~D. {Murtagh}, 76--87, \dodoi{10.1117/12.447161}

\bibitem[{{Gabriel} {et~al.}(2004){Gabriel}, {Denby}, {Fyfe}, {Hoar}, {Ibarra}, {Ojero}, {Osborne}, {Saxton}, {Lammers}, \& {Vacanti}}]{2004ASPC..314..759G}
{Gabriel}, C., {Denby}, M., {Fyfe}, D.~J., {et~al.} 2004, in Astronomical Society of the Pacific Conference Series, Vol. 314, Astronomical Data Analysis Software and Systems (ADASS) XIII, ed. F.~{Ochsenbein}, M.~G. {Allen}, \& D.~{Egret}, 759

\bibitem[{{Ginzburg} \& {Syrovatskii}(1965)}]{1965ARA&A...3..297G}
{Ginzburg}, V.~L., \& {Syrovatskii}, S.~I. 1965, \araa, 3, 297, \dodoi{10.1146/annurev.aa.03.090165.001501}

\bibitem[{{Giommi} {et~al.}(2012){Giommi}, {Polenta}, {L{\"a}hteenm{\"a}ki}, {Thompson}, {Capalbi}, {Cutini}, {Gasparrini}, {Gonz{\'a}lez-Nuevo}, {Le{\'o}n-Tavares}, {L{\'o}pez-Caniego}, {Mazziotta}, {Monte}, {Perri}, {Rain{\`o}}, {Tosti}, {Tramacere}, {Verrecchia}, {Aller}, {Aller}, {Angelakis}, {Bastieri}, {Berdyugin}, {Bonaldi}, {Bonavera}, {Burigana}, {Burrows}, {Buson}, {Cavazzuti}, {Chincarini}, {Colafrancesco}, {Costamante}, {Cuttaia}, {D'Ammando}, {de Zotti}, {Frailis}, {Fuhrmann}, {Galeotta}, {Gargano}, {Gehrels}, {Giglietto}, {Giordano}, {Giroletti}, {Keih{\"a}nen}, {King}, {Krichbaum}, {Lasenby}, {Lavonen}, {Lawrence}, {Leto}, {Lindfors}, {Mandolesi}, {Massardi}, {Max-Moerbeck}, {Michelson}, {Mingaliev}, {Natoli}, {Nestoras}, {Nieppola}, {Nilsson}, {Partridge}, {Pavlidou}, {Pearson}, {Procopio}, {Rachen}, {Readhead}, {Reeves}, {Reimer}, {Reinthal}, {Ricciardi}, {Richards}, {Riquelme}, {Saarinen}, {Sajina}, {Sandri}, {Savolainen}, {Sievers}, {Sillanp{\"a}{\"a}}, {Sotnikova}, {Stevenson},
  {Tagliaferri}, {Takalo}, {Tammi}, {Tavagnacco}, {Terenzi}, {Toffolatti}, {Tornikoski}, {Trigilio}, {Turunen}, {Umana}, {Ungerechts}, {Villa}, {Wu}, {Zacchei}, {Zensus}, \& {Zhou}}]{2012A&A...541A.160G}
{Giommi}, P., {Polenta}, G., {L{\"a}hteenm{\"a}ki}, A., {et~al.} 2012, \aap, 541, A160, \dodoi{10.1051/0004-6361/201117825}

\bibitem[{Heasarc(2014)}]{2014ascl.soft08004N}
Heasarc. 2014, {HEAsoft: Unified Release of FTOOLS and XANADU}, Astrophysics Source Code Library, record ascl:1408.004.
\newblock \doeprint{1408.004}

\bibitem[{{HI4PI Collaboration} {et~al.}(2016){HI4PI Collaboration}, {Ben Bekhti}, {Fl{\"o}er}, {Keller}, {Kerp}, {Lenz}, {Winkel}, {Bailin}, {Calabretta}, {Dedes}, {Ford}, {Gibson}, {Haud}, {Janowiecki}, {Kalberla}, {Lockman}, {McClure-Griffiths}, {Murphy}, {Nakanishi}, {Pisano}, \& {Staveley-Smith}}]{2016A&A...594A.116H}
{HI4PI Collaboration}, {Ben Bekhti}, N., {Fl{\"o}er}, L., {et~al.} 2016, \aap, 594, A116, \dodoi{10.1051/0004-6361/201629178}

\bibitem[{{Holder} {et~al.}(2003){Holder}, {Bond}, {Boyle}, {Bradbury}, {Buckley}, {Carter-Lewis}, {Cui}, {Dowdall}, {Duke}, {de la Calle Perez}, {Falcone}, {Fegan}, {Fegan}, {Finley}, {Fortson}, {Gaidos}, {Gibbs}, {Gammell}, {Hall}, {Hall}, {Hillas}, {Horan}, {Jordan}, {Kertzman}, {Kieda}, {Kildea}, {Knapp}, {Kosack}, {Krawczynski}, {Krennrich}, {LeBohec}, {Linton}, {Lloyd-Evans}, {Moriarty}, {M{\"u}ller}, {Nagai}, {Ong}, {Page}, {Pallassini}, {Petry}, {Power-Mooney}, {Quinn}, {Rebillot}, {Reynolds}, {Rose}, {Schroedter}, {Sembroski}, {Swordy}, {Vassiliev}, {Wakely}, {Walker}, \& {Weekes}}]{2003ApJ...583L...9H}
{Holder}, J., {Bond}, I.~H., {Boyle}, P.~J., {et~al.} 2003, \apjl, 583, L9, \dodoi{10.1086/367816}

\bibitem[{{Hovatta} {et~al.}(2012){Hovatta}, {Lister}, {Aller}, {Aller}, {Homan}, {Kovalev}, {Pushkarev}, \& {Savolainen}}]{2012AJ....144..105H}
{Hovatta}, T., {Lister}, M.~L., {Aller}, M.~F., {et~al.} 2012, \aj, 144, 105, \dodoi{10.1088/0004-6256/144/4/105}

\bibitem[{{Hovatta} {et~al.}(2016){Hovatta}, {Lindfors}, {Blinov}, {Pavlidou}, {Nilsson}, {Kiehlmann}, {Angelakis}, {Fallah Ramazani}, {Liodakis}, {Myserlis}, {Panopoulou}, \& {Pursimo}}]{Hovatta2016}
{Hovatta}, T., {Lindfors}, E., {Blinov}, D., {et~al.} 2016, \aap, 596, A78, \dodoi{10.1051/0004-6361/201628974}

\bibitem[{{Hughes} {et~al.}(1985){Hughes}, {Aller}, \& {Aller}}]{Hughes1985}
{Hughes}, P.~A., {Aller}, H.~D., \& {Aller}, M.~F. 1985, \apj, 298, 301, \dodoi{10.1086/163611}

\bibitem[{{Hughes} \& {Bregman}(2006)}]{Hughes2006}
{Hughes}, P.~A., \& {Bregman}, J.~N., eds. 2006, American Institute of Physics Conference Series, Vol. 856, {Relativistic Jets: The Common Physics of AGN, Microquasars, and Gamma-Ray Bursts}

\bibitem[{{Jorstad} {et~al.}(2005){Jorstad}, {Marscher}, {Lister}, {Stirling}, {Cawthorne}, {Gear}, {G{\'o}mez}, {Stevens}, {Smith}, {Forster}, \& {Robson}}]{Jorstad2005AJ....130.1418J}
{Jorstad}, S.~G., {Marscher}, A.~P., {Lister}, M.~L., {et~al.} 2005, \aj, 130, 1418, \dodoi{10.1086/444593}

\bibitem[{{Jorstad} {et~al.}(2017){Jorstad}, {Marscher}, {Morozova}, {Troitsky}, {Agudo}, {Casadio}, {Foord}, {G{\'o}mez}, {MacDonald}, {Molina}, {L{\"a}hteenm{\"a}ki}, {Tammi}, \& {Tornikoski}}]{jorstad2017}
{Jorstad}, S.~G., {Marscher}, A.~P., {Morozova}, D.~A., {et~al.} 2017, \apj, 846, 98, \dodoi{10.3847/1538-4357/aa8407}

\bibitem[{{Kagan} {et~al.}(2015){Kagan}, {Sironi}, {Cerutti}, \& {Giannios}}]{2015SSRv..191..545K}
{Kagan}, D., {Sironi}, L., {Cerutti}, B., \& {Giannios}, D. 2015, \ssr, 191, 545, \dodoi{10.1007/s11214-014-0132-9}

\bibitem[{{Kapanadze} {et~al.}(2014){Kapanadze}, {Romano}, {Vercellone}, {Kapanadze}, \& {Kharshiladze}}]{Kapanadze2014}
{Kapanadze}, B., {Romano}, P., {Vercellone}, S., {Kapanadze}, S., \& {Kharshiladze}, G. 2014, in Proceedings of Swift: 10 Years of Discovery (SWIFT 10, 142, \dodoi{10.22323/1.233.0142}

\bibitem[{{Katarzy{\'n}ski} {et~al.}(2006){Katarzy{\'n}ski}, {Ghisellini}, {Mastichiadis}, {Tavecchio}, \& {Maraschi}}]{2006A&A...453...47K}
{Katarzy{\'n}ski}, K., {Ghisellini}, G., {Mastichiadis}, A., {Tavecchio}, F., \& {Maraschi}, L. 2006, \aap, 453, 47, \dodoi{10.1051/0004-6361:20054176}

\bibitem[{{Kawabata} {et~al.}(1999){Kawabata}, {Okazaki}, {Akitaya}, {Hirakata}, {Hirata}, {Ikeda}, {Kondoh}, {Masuda}, \& {Seki}}]{Kawabata1999}
{Kawabata}, K.~S., {Okazaki}, A., {Akitaya}, H., {et~al.} 1999, \pasp, 111, 898, \dodoi{10.1086/316387}

\bibitem[{{Kirk} {et~al.}(1996){Kirk}, {Duffy}, \& {Gallant}}]{1996A&A...314.1010K}
{Kirk}, J.~G., {Duffy}, P., \& {Gallant}, Y.~A. 1996, \aap, 314, 1010, \dodoi{10.48550/arXiv.astro-ph/9604056}

\bibitem[{{Krawczynski}(2012)}]{Krawczynski2012}
{Krawczynski}, H. 2012, \apj, 744, 30, \dodoi{10.1088/0004-637X/744/1/30}

\bibitem[{{Krawczynski} {et~al.}(2004){Krawczynski}, {Hughes}, {Horan}, {Aharonian}, {Aller}, {Aller}, {Boltwood}, {Buckley}, {Coppi}, {Fossati}, {G{\"o}tting}, {Holder}, {Horns}, {Kurtanidze}, {Marscher}, {Nikolashvili}, {Remillard}, {Sadun}, \& {Schr{\"o}der}}]{2004ApJ...601..151K}
{Krawczynski}, H., {Hughes}, S.~B., {Horan}, D., {et~al.} 2004, \apj, 601, 151, \dodoi{10.1086/380393}

\bibitem[{{Liodakis} {et~al.}(2019){Liodakis}, {Peirson}, \& {Romani}}]{Liodakis2019}
{Liodakis}, I., {Peirson}, A.~L., \& {Romani}, R.~W. 2019, \apj, 880, 29, \dodoi{10.3847/1538-4357/ab2719}

\bibitem[{{Liodakis} {et~al.}(2022){Liodakis}, {Marscher}, {Agudo}, {Berdyugin}, {Bernardos}, {Bonnoli}, {Borman}, {Casadio}, {Casanova}, {Cavazzuti}, {Rodriguez Cavero}, {Di Gesu}, {Di Lalla}, {Donnarumma}, {Ehlert}, {Errando}, {Escudero}, {Garc{\'\i}a-Comas}, {Ag{\'\i}s-Gonz{\'a}lez}, {Husillos}, {Jormanainen}, {Jorstad}, {Kagitani}, {Kopatskaya}, {Kravtsov}, {Krawczynski}, {Lindfors}, {Larionova}, {Madejski}, {Marin}, {Marchini}, {Marshall}, {Morozova}, {Massaro}, {Masiero}, {Mawet}, {Middei}, {Millar-Blanchaer}, {Myserlis}, {Negro}, {Nilsson}, {O'Dell}, {Omodei}, {Pacciani}, {Paggi}, {Panopoulou}, {Peirson}, {Perri}, {Petrucci}, {Poutanen}, {Puccetti}, {Romani}, {Sakanoi}, {Savchenko}, {Sota}, {Tavecchio}, {Tinyanont}, {Vasilyev}, {Weaver}, {Zhovtan}, {Antonelli}, {Bachetti}, {Baldini}, {Baumgartner}, {Bellazzini}, {Bianchi}, {Bongiorno}, {Bonino}, {Brez}, {Bucciantini}, {Capitanio}, {Castellano}, {Ciprini}, {Costa}, {De Rosa}, {Del Monte}, {Di Marco}, {Doroshenko}, {Dov{\v{c}}iak}, {Enoto},
  {Evangelista}, {Fabiani}, {Ferrazzoli}, {Garcia}, {Gunji}, {Hayashida}, {Heyl}, {Iwakiri}, {Karas}, {Kitaguchi}, {Kolodziejczak}, {La Monaca}, {Latronico}, {Maldera}, {Manfreda}, {Marinucci}, {Matt}, {Mitsuishi}, {Mizuno}, {Muleri}, {Ng}, {Oppedisano}, {Papitto}, {Pavlov}, {Pesce-Rollins}, {Pilia}, {Possenti}, {Ramsey}, {Rankin}, {Ratheesh}, {Sgr{\'o}}, {Slane}, {Soffitta}, {Spandre}, {Tamagawa}, {Taverna}, {Tawara}, {Tennant}, {Thomas}, {Tombesi}, {Trois}, {Tsygankov}, {Turolla}, {Vink}, {Weisskopf}, {Wu}, {Xie}, \& {Zane}}]{Liodakis2022-Mrk501}
{Liodakis}, I., {Marscher}, A.~P., {Agudo}, I., {et~al.} 2022, \nat, 611, 677, \dodoi{10.1038/s41586-022-05338-0}

\bibitem[{Lyubarsky(2010)}]{Lyubarsky2010}
Lyubarsky, Y.~E. 2010, Monthly Notices of the Royal Astronomical Society, 402, 353, \dodoi{10.1111/j.1365-2966.2009.15877.x}

\bibitem[{{Madsen} {et~al.}(2017){Madsen}, {Beardmore}, {Forster}, {Guainazzi}, {Marshall}, {Miller}, {Page}, \& {Stuhlinger}}]{2017AJ....153....2M}
{Madsen}, K.~K., {Beardmore}, A.~P., {Forster}, K., {et~al.} 2017, \aj, 153, 2, \dodoi{10.3847/1538-3881/153/1/2}

\bibitem[{{Marrone} \& {Rao}(2008)}]{Marrone2008}
{Marrone}, D.~P., \& {Rao}, R. 2008, in \procspie, Vol. 7020, Millimeter and Submillimeter Detectors and Instrumentation for Astronomy IV, ed. W.~D. {Duncan}, W.~S. {Holland}, S.~{Withington}, \& J.~{Zmuidzinas}, 70202B, \dodoi{10.1117/12.788677}

\bibitem[{{Marscher}(2014)}]{2014ApJ...780...87M}
{Marscher}, A.~P. 2014, \apj, 780, 87, \dodoi{10.1088/0004-637X/780/1/87}

\bibitem[{{Marscher} \& {Jorstad}(2021)}]{marscher2021}
{Marscher}, A.~P., \& {Jorstad}, S.~G. 2021, Galaxies, 9, 27, \dodoi{10.3390/galaxies9020027}

\bibitem[{{Marscher} \& {Jorstad}(2022)}]{marscher2022}
---. 2022, Universe, 8, 644, \dodoi{10.3390/universe8120644}

\bibitem[{{Middei} {et~al.}(2023{\natexlab{a}}){Middei}, {Perri}, {Puccetti}, {Liodakis}, {Di Gesu}, {Marscher}, {Rodriguez Cavero}, {Tavecchio}, {Donnarumma}, {Laurenti}, {Jorstad}, {Agudo}, {Marshall}, {Pacciani}, {Kim}, {Aceituno}, {Bonnoli}, {Casanova}, {Ag{\'\i}s-Gonz{\'a}lez}, {Sota}, {Casadio}, {Escudero}, {Myserlis}, {Sievers}, {Kouch}, {Lindfors}, {Gurwell}, {Keating}, {Rao}, {Kang}, {Lee}, {Kim}, {Cheong}, {Jeong}, {Angelakis}, {Kraus}, {Antonelli}, {Bachetti}, {Baldini}, {Baumgartner}, {Bellazzini}, {Bianchi}, {Bongiorno}, {Bonino}, {Brez}, {Bucciantini}, {Capitanio}, {Castellano}, {Cavazzuti}, {Chen}, {Ciprini}, {Costa}, {De Rosa}, {Del Monte}, {Di Lalla}, {Di Marco}, {Doroshenko}, {Dov{\v{c}}iak}, {Ehlert}, {Enoto}, {Evangelista}, {Fabiani}, {Ferrazzoli}, {Garc{\'\i}a}, {Gunji}, {Hayashida}, {Heyl}, {Iwakiri}, {Kaaret}, {Karas}, {Kislat}, {Kitaguchi}, {Kolodziejczak}, {Krawczynski}, {La Monaca}, {Latronico}, {Maldera}, {Manfreda}, {Marin}, {Marinucci}, {Massaro}, {Matt}, {Mitsuishi}, {Mizuno},
  {Muleri}, {Negro}, {Ng}, {O'Dell}, {Omodei}, {Oppedisano}, {Papitto}, {Pavlov}, {Peirson}, {Pesce-Rollins}, {Petrucci}, {Pilia}, {Possenti}, {Poutanen}, {Ramsey}, {Rankin}, {Ratheesh}, {Roberts}, {Romani}, {Sgr{\`o}}, {Slane}, {Soffitta}, {Spandre}, {Swartz}, {Tamagawa}, {Taverna}, {Tawara}, {Tennant}, {Thomas}, {Tombesi}, {Trois}, {Tsygankov}, {Turolla}, {Vink}, {Weisskopf}, {Wu}, {Xie}, \& {Zane}}]{ixpe1553}
{Middei}, R., {Perri}, M., {Puccetti}, S., {et~al.} 2023{\natexlab{a}}, \apjl, 953, L28, \dodoi{10.3847/2041-8213/acec3e}

\bibitem[{{Middei} {et~al.}(2023{\natexlab{b}}){Middei}, {Liodakis}, {Perri}, {Puccetti}, {Cavazzuti}, {Di Gesu}, {Ehlert}, {Madejski}, {Marscher}, {Marshall}, {Muleri}, {Negro}, {Jorstad}, {Ag{\'\i}s-Gonz{\'a}lez}, {Agudo}, {Bonnoli}, {Bernardos}, {Casanova}, {Garc{\'\i}a-Comas}, {Husillos}, {Marchini}, {Sota}, {Kouch}, {Lindfors}, {Borman}, {Kopatskaya}, {Larionova}, {Morozova}, {Savchenko}, {Vasilyev}, {Zhovtan}, {Casadio}, {Escudero}, {Myserlis}, {Hales}, {Kameno}, {Kneissl}, {Messias}, {Nagai}, {Blinov}, {Bourbah}, {Kiehlmann}, {Kontopodis}, {Mandarakas}, {Romanopoulos}, {Skalidis}, {Vervelaki}, {Masiero}, {Mawet}, {Millar-Blanchaer}, {Panopoulou}, {Tinyanont}, {Berdyugin}, {Kagitani}, {Kravtsov}, {Sakanoi}, {Imazawa}, {Sasada}, {Fukazawa}, {Kawabata}, {Uemura}, {Mizuno}, {Nakaoka}, {Akitaya}, {Gurwell}, {Rao}, {Di Lalla}, {Cibrario}, {Donnarumma}, {Kim}, {Omodei}, {Pacciani}, {Poutanen}, {Tavecchio}, {Antonelli}, {Bachetti}, {Baldini}, {Baumgartner}, {Bellazzini}, {Bianchi}, {Bongiorno}, {Bonino},
  {Brez}, {Bucciantini}, {Capitanio}, {Castellano}, {Ciprini}, {Costa}, {De Rosa}, {Del Monte}, {Di Marco}, {Doroshenko}, {Dov{\v{c}}iak}, {Enoto}, {Evangelista}, {Fabiani}, {Ferrazzoli}, {Garcia}, {Gunji}, {Hayashida}, {Heyl}, {Iwakiri}, {Karas}, {Kitaguchi}, {Kolodziejczak}, {Krawczynski}, {La Monaca}, {Latronico}, {Maldera}, {Manfreda}, {Marin}, {Marinucci}, {Massaro}, {Matt}, {Mitsuishi}, {Ng}, {O'Dell}, {Oppedisano}, {Papitto}, {Pavlov}, {Peirson}, {Pesce-Rollins}, {Petrucci}, {Pilia}, {Possenti}, {Ramsey}, {Rankin}, {Ratheesh}, {Romani}, {Sgr{\'o}}, {Slane}, {Soffitta}, {Spandre}, {Tamagawa}, {Taverna}, {Tawara}, {Tennant}, {Thomas}, {Tombesi}, {Trois}, {Tsygankov}, {Turolla}, {Vink}, {Weisskopf}, {Wu}, {Xie}, \& {Zane}}]{Middei2023}
{Middei}, R., {Liodakis}, I., {Perri}, M., {et~al.} 2023{\natexlab{b}}, \apjl, 942, L10, \dodoi{10.3847/2041-8213/aca281}

\bibitem[{{Moffat}(1969)}]{Moffat1969}
{Moffat}, A.~F.~J. 1969, \aap, 3, 455

\bibitem[{{Nilsson} {et~al.}(2007){Nilsson}, {Pasanen}, {Takalo}, {Lindfors}, {Berdyugin}, {Ciprini}, \& {Pforr}}]{Nilsson2007}
{Nilsson}, K., {Pasanen}, M., {Takalo}, L.~O., {et~al.} 2007, \aap, 475, 199, \dodoi{10.1051/0004-6361:20077624}

\bibitem[{{Nilsson} {et~al.}(2018){Nilsson}, {Lindfors}, {Takalo}, {Reinthal}, {Berdyugin}, {Sillanp{\"a}{\"a}}, {Ciprini}, {Halkola}, {Hein{\"a}m{\"a}ki}, {Hovatta}, {Kadenius}, {Nurmi}, {Ostorero}, {Pasanen}, {Rekola}, {Saarinen}, {Sainio}, {Tuominen}, {Villforth}, {Vornanen}, \& {Zaprudin}}]{Nilsson2018}
{Nilsson}, K., {Lindfors}, E., {Takalo}, L.~O., {et~al.} 2018, \aap, 620, A185, \dodoi{10.1051/0004-6361/201833621}

\bibitem[{{Nishikawa} {et~al.}(2003){Nishikawa}, {Hardee}, {Richardson}, {Preece}, {Sol}, \& {Fishman}}]{2003ApJ...595..555N}
{Nishikawa}, K.~I., {Hardee}, P., {Richardson}, G., {et~al.} 2003, \apj, 595, 555, \dodoi{10.1086/377260}

\bibitem[{{Panopoulou} {et~al.}(2015){Panopoulou}, {Tassis}, {Blinov}, {Pavlidou}, {King}, {Paleologou}, {Ramaprakash}, {Angelakis}, {Balokovi{\'c}}, {Das}, {Feiler}, {Hovatta}, {Khodade}, {Kiehlmann}, {Kus}, {Kylafis}, {Liodakis}, {Mahabal}, {Modi}, {Myserlis}, {Papadakis}, {Papamastorakis}, {Pazderska}, {Pazderski}, {Pearson}, {Rajarshi}, {Readhead}, {Reig}, \& {Zensus}}]{Panopoulou2015}
{Panopoulou}, G., {Tassis}, K., {Blinov}, D., {et~al.} 2015, \mnras, 452, 715, \dodoi{10.1093/mnras/stv1301}

\bibitem[{{Perlman} {et~al.}(1996){Perlman}, {Stocke}, {Schachter}, {Elvis}, {Ellingson}, {Urry}, {Potter}, {Impey}, \& {Kolchinsky}}]{1996ApJS..104..251P}
{Perlman}, E.~S., {Stocke}, J.~T., {Schachter}, J.~F., {et~al.} 1996, \apjs, 104, 251, \dodoi{10.1086/192300}

\bibitem[{{Piner} {et~al.}(2010){Piner}, {Pant}, \& {Edwards}}]{2010ApJ...723.1150P}
{Piner}, B.~G., {Pant}, N., \& {Edwards}, P.~G. 2010, \apj, 723, 1150, \dodoi{10.1088/0004-637X/723/2/1150}

\bibitem[{{Primiani} {et~al.}(2016){Primiani}, {Young}, {Young}, {Patel}, {Wilson}, {Vertatschitsch}, {Chitwood}, {Srinivasan}, {MacMahon}, \& {Weintroub}}]{Primiani2016}
{Primiani}, R.~A., {Young}, K.~H., {Young}, A., {et~al.} 2016, Journal of Astronomical Instrumentation, 5, 1641006, \dodoi{10.1142/S2251171716410063}

\bibitem[{{Ramaprakash} {et~al.}(2019){Ramaprakash}, {Rajarshi}, {Das}, {Khodade}, {Modi}, {Panopoulou}, {Maharana}, {Blinov}, {Angelakis}, {Casadio}, {Fuhrmann}, {Hovatta}, {Kiehlmann}, {King}, {Kylafis}, {Kougentakis}, {Kus}, {Mahabal}, {Marecki}, {Myserlis}, {Paterakis}, {Paleologou}, {Liodakis}, {Papadakis}, {Papamastorakis}, {Pavlidou}, {Pazderski}, {Pearson}, {Readhead}, {Reig}, {S{\l}owikowska}, {Tassis}, \& {Zensus}}]{Ramaprakash2019}
{Ramaprakash}, A.~N., {Rajarshi}, C.~V., {Das}, H.~K., {et~al.} 2019, \mnras, 485, 2355, \dodoi{10.1093/mnras/stz557}

\bibitem[{{Romanova} \& {Lovelace}(1992)}]{1992A&A...262...26R}
{Romanova}, M.~M., \& {Lovelace}, R.~V.~E. 1992, \aap, 262, 26

\bibitem[{{Rybicki} \& {Lightman}(1979)}]{1979rpa..book.....R}
{Rybicki}, G.~B., \& {Lightman}, A.~P. 1979, {Radiative processes in astrophysics}

\bibitem[{{S{\'e}rsic}(1963)}]{Sersic1963}
{S{\'e}rsic}, J.~L. 1963, Boletin de la Asociacion Argentina de Astronomia La Plata Argentina, 6, 41

\bibitem[{{Sersic}(1968)}]{Sersic1968}
{Sersic}, J.~L. 1968, {Atlas de Galaxias Australes}

\bibitem[{{Sironi} {et~al.}(2015{\natexlab{a}}){Sironi}, {Keshet}, \& {Lemoine}}]{2015SSRv..191..519S}
{Sironi}, L., {Keshet}, U., \& {Lemoine}, M. 2015{\natexlab{a}}, \ssr, 191, 519, \dodoi{10.1007/s11214-015-0181-8}

\bibitem[{{Sironi} {et~al.}(2015{\natexlab{b}}){Sironi}, {Petropoulou}, \& {Giannios}}]{Sironi2015}
{Sironi}, L., {Petropoulou}, M., \& {Giannios}, D. 2015{\natexlab{b}}, \mnras, 450, 183, \dodoi{10.1093/mnras/stv641}

\bibitem[{{Sironi} \& {Spitkovsky}(2014)}]{2014ApJ...783L..21S}
{Sironi}, L., \& {Spitkovsky}, A. 2014, \apjl, 783, L21, \dodoi{10.1088/2041-8205/783/1/L21}

\bibitem[{{Smith} {et~al.}(2007){Smith}, {Williams}, {Schmidt}, {Diamond-Stanic}, \& {Means}}]{Smith2007ApJ...663..118}
{Smith}, P.~S., {Williams}, G.~G., {Schmidt}, G.~D., {Diamond-Stanic}, A.~M., \& {Means}, D.~L. 2007, \apj, 663, 118, \dodoi{10.1086/517992}

\bibitem[{{Stroh} \& {Falcone}(2013)}]{Falcone2013ApJS..207...28S}
{Stroh}, M.~C., \& {Falcone}, A.~D. 2013, \apjs, 207, 28, \dodoi{10.1088/0067-0049/207/2/28}

\bibitem[{{Str{\"u}der} {et~al.}(2001){Str{\"u}der}, {Briel}, {Dennerl}, {Hartmann}, {Kendziorra}, {Meidinger}, {Pfeffermann}, {Reppin}, {Aschenbach}, {Bornemann}, {Br{\"a}uninger}, {Burkert}, {Elender}, {Freyberg}, {Haberl}, {Hartner}, {Heuschmann}, {Hippmann}, {Kastelic}, {Kemmer}, {Kettenring}, {Kink}, {Krause}, {M{\"u}ller}, {Oppitz}, {Pietsch}, {Popp}, {Predehl}, {Read}, {Stephan}, {St{\"o}tter}, {Tr{\"u}mper}, {Holl}, {Kemmer}, {Soltau}, {St{\"o}tter}, {Weber}, {Weichert}, {von Zanthier}, {Carathanassis}, {Lutz}, {Richter}, {Solc}, {B{\"o}ttcher}, {Kuster}, {Staubert}, {Abbey}, {Holland}, {Turner}, {Balasini}, {Bignami}, {La Palombara}, {Villa}, {Buttler}, {Gianini}, {Lain{\'e}}, {Lumb}, \& {Dhez}}]{2001A&A...365L..18S}
{Str{\"u}der}, L., {Briel}, U., {Dennerl}, K., {et~al.} 2001, \aap, 365, L18, \dodoi{10.1051/0004-6361:20000066}

\bibitem[{{Tagliaferri} {et~al.}(2008){Tagliaferri}, {Foschini}, {Ghisellini}, {Maraschi}, {pre=''and'' post=''}, {and '' affil=''4''>G. Tosti}, {Albert}, {Aliu}, {Anderhub}, {Antoranz}, {Baixeras}, {Barrio}, {Bartko}, {Bastieri}, {Becker}, {Bednarek}, {Bedyugin}, {Berger}, {Bigongiari}, {Biland}, {Bock}, {Bordas}, {Bosch-Ramon}, {Bretz}, {Britvitch}, {Camara}, {Carmona}, {Chilingarian}, {Coarasa}, {Commichau}, {Contreras}, {Cortina}, {Costado}, {Curtef}, {Danielyan}, {Dazzi}, {De Angelis}, {Delgado}, {de los Reyes}, {De Lotto}, {Dorner}, {Doro}, {Errando}, {Fagiolini}, {Ferenc}, {Fern{\'a}ndez}, {Firpo}, {Fonseca}, {Font}, {Fuchs}, {Galante}, {Garc{\'\i}a-L{\'o}pez}, {Garczarczyk}, {Gaug}, {Giller}, {Goebel}, {Hakobyan}, {Hayashida}, {Hengstebeck}, {Herrero}, {H{\"o}hne}, {Hose}, {Huber}, {Hsu}, {Jacon}, {Jogler}, {Kosyra}, {Kranich}, {Kritzer}, {Laille}, {Lindfors}, {Lombardi}, {Longo}, {L{\'o}pez}, {Lorenz}, {Majumdar}, {Maneva}, {Mannheim}, {Mariotti}, {Mart{\'\i}nez}, {Mazin}, {Merck}, {Meucci}, {Meyer},
  {Miranda}, {Mirzoyan}, {Mizobuchi}, {Moralejo}, {Nieto}, {Nilsson}, {Ninkovic}, {O{\~n}a-Wilhelmi}, {Otte}, {Oya}, {Panniello}, {Paoletti}, {Paredes}, {Pasanen}, {Pascoli}, {Pauss}, {Pegna}, {Persic}, {Peruzzo}, {Piccioli}, {Prandini}, {Puchades}, {Raymers}, {Rhode}, {Rib{\'o}}, {Rico}, {Rissi}, {Robert}, {R{\"u}gamer}, {Saggion}, {Saito}, {S{\'a}nchez}, {Sartori}, {Scalzotto}, {Scapin}, {Schmitt}, {Schweizer}, {Shayduk}, {Shinozaki}, {Shore}, {Sidro}, {Sillanp{\"a}{\"a}}, {Sobczynska}, {Spanier}, {Stamerra}, {Stark}, {Takalo}, {Tavecchio}, {Temnikov}, {Tescaro}, {Teshima}, {Torres}, {Turini}, {Vankov}, {Venturini}, {Vitale}, {Wagner}, {Wibig}, {Wittek}, {Zandanel}, {Zanin}, {Zapatero}, \& {MAGIC Collaboration''}}]{2008ApJ...679.1029T}
{Tagliaferri}, G., {Foschini}, L., {Ghisellini}, G., {et~al.} 2008, \apj, 679, 1029, \dodoi{10.1086/586731}

\bibitem[{{Tavecchio}(2021)}]{tavecchio2021}
{Tavecchio}, F. 2021, Galaxies, 9, 37, \dodoi{10.3390/galaxies9020037}

\bibitem[{{Tavecchio} {et~al.}(2018){Tavecchio}, {Landoni}, {Sironi}, \& {Coppi}}]{tavecchio2018}
{Tavecchio}, F., {Landoni}, M., {Sironi}, L., \& {Coppi}, P. 2018, \mnras, 480, 2872, \dodoi{10.1093/mnras/sty1491}

\bibitem[{{Tavecchio} {et~al.}(2020){Tavecchio}, {Landoni}, {Sironi}, \& {Coppi}}]{tavecchio2020}
---. 2020, \mnras, 498, 599, \dodoi{10.1093/mnras/staa2457}

\bibitem[{{Tchekhovskoy} {et~al.}(2009){Tchekhovskoy}, {McKinney}, \& {Narayan}}]{2009ApJ...699.1789T}
{Tchekhovskoy}, A., {McKinney}, J.~C., \& {Narayan}, R. 2009, \apj, 699, 1789, \dodoi{10.1088/0004-637X/699/2/1789}

\bibitem[{{Thum} {et~al.}(2018){Thum}, {Agudo}, {Molina}, {Casadio}, {G{\'o}mez}, {Morris}, {Ramakrishnan}, \& {Sievers}}]{Thum2018}
{Thum}, C., {Agudo}, I., {Molina}, S.~N., {et~al.} 2018, \mnras, 473, 2506, \dodoi{10.1093/mnras/stx2436}

\bibitem[{{Trujillo} {et~al.}(2001){Trujillo}, {Aguerri}, {Cepa}, \& {Guti{\'e}rrez}}]{Trujillo2001}
{Trujillo}, I., {Aguerri}, J.~A.~L., {Cepa}, J., \& {Guti{\'e}rrez}, C.~M. 2001, \mnras, 328, 977, \dodoi{10.1046/j.1365-8711.2001.04937.x}

\bibitem[{{Urry} \& {Padovani}(1995)}]{1995PASP..107..803U}
{Urry}, C.~M., \& {Padovani}, P. 1995, \pasp, 107, 803, \dodoi{10.1086/133630}

\bibitem[{{Vlahakis} \& {K{\"o}nigl}(2004)}]{Vlahakis2004}
{Vlahakis}, N., \& {K{\"o}nigl}, A. 2004, \apj, 605, 656, \dodoi{10.1086/382670}

\bibitem[{{Weaver} {et~al.}(2022){Weaver}, {Jorstad}, {Marscher}, {Morozova}, {Troitsky}, {Agudo}, {G{\'o}mez}, {L{\"a}hteenm{\"a}ki}, {Tammi}, \& {Tornikoski}}]{weaver2022}
{Weaver}, Z.~R., {Jorstad}, S.~G., {Marscher}, A.~P., {et~al.} 2022, \apjs, 260, 12, \dodoi{10.3847/1538-4365/ac589c}

\bibitem[{{Weisskopf} {et~al.}(2022){Weisskopf}, {Soffitta}, {Baldini}, {Ramsey}, {O'Dell}, {Romani}, {Matt}, {Deininger}, {Baumgartner}, {Bellazzini}, {Costa}, {Kolodziejczak}, {Latronico}, {Marshall}, {Muleri}, {Bongiorno}, {Tennant}, {Bucciantini}, {Dovciak}, {Marin}, {Marscher}, {Poutanen}, {Slane}, {Turolla}, {Kalinowski}, {Di Marco}, {Fabiani}, {Minuti}, {La Monaca}, {Pinchera}, {Rankin}, {Sgro'}, {Trois}, {Xie}, {Alexander}, {Allen}, {Amici}, {Andersen}, {Antonelli}, {Antoniak}, {Attin{\`a}}, {Barbanera}, {Bachetti}, {Baggett}, {Bladt}, {Brez}, {Bonino}, {Boree}, {Borotto}, {Breeding}, {Brienza}, {Bygott}, {Caporale}, {Cardelli}, {Carpentiero}, {Castellano}, {Castronuovo}, {Cavalli}, {Cavazzuti}, {Ceccanti}, {Centrone}, {Citraro}, {D'Amico}, {D'Alba}, {Di Gesu}, {Del Monte}, {Dietz}, {Di Lalla}, {Persio}, {Dolan}, {Donnarumma}, {Evangelista}, {Ferrant}, {Ferrazzoli}, {Ferrie}, {Footdale}, {Forsyth}, {Foster}, {Garelick}, {Gunji}, {Gurnee}, {Head}, {Hibbard}, {Johnson}, {Kelly}, {Kilaru}, {Lefevre},
  {Roy}, {Loffredo}, {Lorenzi}, {Lucchesi}, {Maddox}, {Magazzu}, {Maldera}, {Manfreda}, {Mangraviti}, {Marengo}, {Marrocchesi}, {Massaro}, {Mauger}, {McCracken}, {McEachen}, {Mize}, {Mereu}, {Mitchell}, {Mitsuishi}, {Morbidini}, {Mosti}, {Nasimi}, {Negri}, {Negro}, {Nguyen}, {Nitschke}, {Nuti}, {Onizuka}, {Oppedisano}, {Orsini}, {Osborne}, {Pacheco}, {Paggi}, {Painter}, {Pavelitz}, {Pentz}, {Piazzolla}, {Perri}, {Pesce-Rollins}, {Peterson}, {Pilia}, {Profeti}, {Puccetti}, {Ranganathan}, {Ratheesh}, {Reedy}, {Root}, {Rubini}, {Ruswick}, {Sanchez}, {Sarra}, {Santoli}, {Scalise}, {Sciortino}, {Schroeder}, {Seek}, {Sosdian}, {Spandre}, {Speegle}, {Tamagawa}, {Tardiola}, {Tobia}, {Thomas}, {Valerie}, {Vimercati}, {Walden}, {Weddendorf}, {Wedmore}, {Welch}, {Zanetti}, \& {Zanetti}}]{weisskopf2022}
{Weisskopf}, M.~C., {Soffitta}, P., {Baldini}, L., {et~al.} 2022, Journal of Astronomical Telescopes, Instruments, and Systems, 8, 026002, \dodoi{10.1117/1.JATIS.8.2.026002}

\bibitem[{{Werner} {et~al.}(2018){Werner}, {Uzdensky}, {Begelman}, {Cerutti}, \& {Nalewajko}}]{2018MNRAS.473.4840W}
{Werner}, G.~R., {Uzdensky}, D.~A., {Begelman}, M.~C., {Cerutti}, B., \& {Nalewajko}, K. 2018, \mnras, 473, 4840, \dodoi{10.1093/mnras/stx2530}

\bibitem[{{Zhang} {et~al.}(2019){Zhang}, {Fang}, {Li}, {Giannios}, {B{\"o}ttcher}, \& {Buson}}]{zhang2019}
{Zhang}, H., {Fang}, K., {Li}, H., {et~al.} 2019, \apj, 876, 109, \dodoi{10.3847/1538-4357/ab158d}

\bibitem[{{Zhang} {et~al.}(2022){Zhang}, {Li}, {Giannios}, {Guo}, {Thiersen}, {B{\"o}ttcher}, {Lewis}, \& {Venters}}]{zhang2022}
{Zhang}, H., {Li}, X., {Giannios}, D., {et~al.} 2022, \apj, 924, 90, \dodoi{10.3847/1538-4357/ac3669}

\bibitem[{{Zhang} {et~al.}(2023){Zhang}, {Marscher}, {Guo}, {Giannios}, {Li}, \& {Negro}}]{2023arXiv230113316Z}
{Zhang}, H., {Marscher}, A.~P., {Guo}, F., {et~al.} 2023, \apj, 949, 71, \dodoi{10.3847/1538-4357/acc657}

\end{thebibliography}
\bibliographystyle{aasjournal}

\end{document}